\documentclass[12pt,a4paper]{article}
\usepackage[cp1250]{inputenc}
\begin{document}
\title{Symmetry-reduced Loop Quantum Gravity: Plane Waves, Flat Space and the Hamiltonian Constraint}
\author{F. Hinterleitner\footnote{e-mail: franz@physics.muni.cz}\\ Institute of Theoretical Physics and Astrophysics,
Masaryk University,\\ Kotl\'{a}\v{r}sk\'a 2, 611 37 Brno, Czech
Republic} \maketitle
\begin{abstract}
Loop quantum gravity (LQG) methods are applied to a symmetry-reduced
model with homogeneity in two dimensions, derived from a Gowdy
model. The conditions for the propagation of unidirectional plane
gravitational waves at exactly the speed of light are set up in the
form of two null Killing equations in terms of Ashtekar variables
and imposed as operators on the quantum states of the system.

Owing to symmetry reduction the gauge group of the system reduces
formally from $SU(2)$ to $U(1)$. Under the assumption of equal
spacing of the holonomy eigenvalues, the solutions are not
normalizable in the sense of the usual inner product on $U(1)$.
Taking over the inner product from the genuine gauge group $SU(2)$
of LQG renders the obtained states normalizable, nevertheless
fluctuations of geometrical quantities remain divergent. In
consequence, the solutions of the (non-commuting) Killing conditions
have to be renormalized. Two kinds of renormalization are presented.
The combination of the occurrence of non-commuting Killing operators
and the necessity of renormalization indicates fluctuations of the
propagation speed, i.\,e. dispersion of gravitational waves.

Finally the same methods are applied to the Hamiltonian constraint
with the same result concerning normalizability. After
renormalization the constraint is not exactly satisfied any more,
which suggests the presence of some kind of interacting matter.
\end{abstract}

\section{Introduction}
The main motivation for this study is to investigate the occurrence
of quantum vacuum dispersion of electromagnetic or gravitational
waves. Due to the discreteness of space at the Planck scale in LQG
the speculation arises that the speed of extremely high-energy
photons might depart from the constant value $c$. There is extensive
literature on this topic, energy-dependent speed of light, Lorentz
invariance violations, modifications of special relativity, etc.
(see an overview in \cite{Florian}), most of which are based on
semiclassical models. A more specific example, dealing with
gravitational wave dispersion, is \cite{Hos}, where the effects of
inhomogeneous perturbations of homogenous cosmological models on
gravitational and electromagnetic waves are studied. In the present
paper we study only the propagation and dispersion of plane
gravitational waves as background for Einstein-Maxwell waves, the
latter ones to be studied at a later stage. We set up a simple, but
fully quantized model, symmetry reduced by homogeneity in two
dimensions, but manifestly inhomogeneous in the third one, so that
the dynamics happens in one spatial dimension. This restriction to
effectively 1+1 dimensions is very convenient for the study of the
speed of waves, it enables us to obtain results from a fully
quantized model, where unidirectional waves are singled out by the
action of ``unidirectionality operators" on more general states.

A few words to Loop Quantum Gravity in general: It is an attempt to
formulate General Relativity (GR) as a Yang-Mils like theory with
the gauge group $SU(2)$. In the canonical version
\cite{AA,Carlo,Bohr,Dah,bod} (there is also a covariant version
\cite{Rov}) spatial components of an (independent) connection play
the role of a gauge potential. The spatial metric is replaced by
orthonormal triads at every point in space,
$e_i={e^a}_i(x)\,\partial_a$ with the triad matrix ${e^a}_i$. The
canonical variables, conjugate to the connection are densitized
triads ${E^a}_i=e\,{e^a}_i$, where $e$ is the determinant of the
inverse triad matrix.

The dynamics of the system is determined by a Hamiltonian, which
consists of constraints, corresponding to the general covariance
(coordinate independence) of the theory so that the dynamical
evolution is merely a form of gauge (for constraints and gauge
freedom see e.\,g. \cite{HT}). The constraints are the Hamiltonian
constraint, generating the evolution from one spacelike hypersurface
to the next one, the diffeo constraint, generating spatial
diffeomorphisms, and the Gau\ss\ constraint, generating triad
rotations.

In order to keep the constraints first class, the connection is
assumed as only partially independent from the triads (or,
equivalently, from the metric), an approach between Einstein-Hilbert
and Hilbert-Palatini. The independent part is the extrinsic
curvature of spacelike hypersurfaces, $K_{ab}$. The usual connection
in LQG is the Ashtekar connection
${A_a}^i={\Gamma_a}^i-\beta{K_a}^i$, where ${\Gamma_a}^i$ is the
spin connection derived from the triad (Levi-Civita in an orthogonal
basis), ${K_a}^i$ is the extrinsic curvature with one space-time
index $a$ and one orthogonal index $i$ (see (\ref{ext}) and
(\ref{K}) below), and $\beta$ is the Barbero-Immirzi parameter, a
real number.

In some simplified models \cite{MB,BS} like the subject of this
paper, or the underlying Gowdy model in \cite{BD1,BD2}, ${K_a}^i$
alone instead of ${A_a}^i$ is widely used as conjugate variable to
${E^a}_i$, as the Poisson brackets of ${A_a}^i$ and ${K_a}^i$ with
${E^b}_j$ are the same, due to the dependence of $\Gamma$ on $E$.
$E$, sometimes called the ``electric field of gravity" gives rise to
an operator of area.

Quantum states in full LQG are so-called spin networks (SNW) in
3-dimensional space, made from edges and vertices, with the former
ones equipped with irreducible representations of $SU(2)$ in form of
holonomies, and carrying quanta of area, and the latter ones
carrying quanta of volume. Here, owing to the effective
one-dimensionality of the considered model, SNWs will be
one-dimensional sequences of edges and vertices, and holonomies
appear as simple phase factors. In correspondence, state functions
may be considered as $U(1)$ elements - the usual kind of holonomies
along the edges, and vertex functions ( = ``point holonomies") to be
defined at the vertices. The latter ones replace edge holonomies in
the homogeneous directions, transversal to wave propagation. Their
meaning is the following: In this model edge holonomies characterize
only the varying physical area of transversal surfaces with a
certain fiducial coordinate area, the quadrupole nature of the
polarized gravitational waves is encoded in two point holonomies at
every vertex.

Quantum operators: The connection acts in form of holonomy
operators, raising or lowering the order of the $SU(2)$
representation at an edge. $E$ acts as conjugate derivative
operator, its eigenvalues are area eigenvalues. The main
simplification, to be applied among others in the following, is the
reduction of the gauge group from $SU(2)$ to $U(1)$, due to the
homogeneity of the model in the transversal directions of a
gravitational wave.

Other approaches to the quantization of plane gravitational waves
with LQG methods can be found, for example, in \cite{DN1,DN2}, for a
classical canonical formulation see e.\,g \cite{AB}.

Concretely, in Section 2 the dispersion-free propagation of
classical plane gravitational waves is briefly introduced on the
basis of a polarized Gowdy model outlined above and translated into
the existence of a null Killing vector field in the direction of
propagation, leading to two independent conditions.

In Section 3 the system is quantized according to the rules of loop
quantum gravity by defining triad and connection operators, adapted
to the above-mentioned symmetry reductions.

In the main part of the paper, beginning with Section 4, the Killing
conditions are formulated in terms of loop quantum operators acting
locally on quantum states on one-dimensional spin networks,
effectively on the vertex functions, whereas the edge functions
remain undetermined. The formulation of the Killing conditions,
containing derivatives as discrete operators, is ambiguous. Two
versions are considered, one where the operators commute at every
vertex, but not at neighboring vertices, and a second one with
non-commuting operators at one and the same vertex. So, in any case,
both Killing conditions cannot be simultaneously satisfied exactly
in quantum theory.

The solutions are not normalizable, or at least not in an apparently
obvious way with the measure in $U(1)$. Even after making the states
normalizable by choosing a more convenient inner product, taken over
from $SU(2)$, a further modification, applied in Section 5, is
necessary to render also length and volume fluctuations finite.
These modifications violate the Killing conditions and the resulting
departure from uniform propagation at exactly the speed of light is
negatively correlated to the uncertainties of metrical quantities.

In Section 6 the methods developed in the main part lead to
solutions of a certain discrete version of the Hamiltonian
constraint, constructed in analogy to the Killing conditions.
Similar to the solutions of the latter, they require
renormalization.

\section{Classical background}
\subsection{Unidirectional gravitational waves}
To establish a quantum model for unidirectional plane gravitational
waves, we start with a rather restricted sector of full General
Relativity, the polarized cosmological Gowdy model \cite{G} in a
quantized version \cite{BD1,BD2} and impose further restrictions to
unidirectionality at the classical level: As in space-times with
unidirectional plane waves there exists always an isometry in the
null direction along the wave, we postulate a null Killing field in
form of conditions on the phase space variables.

For the sake of a better intuition we begin with an even more
restricted solution that describes precisely unidirectional plane
waves, according to Ehlers and Kundt \cite{EK}, with the following
space-time metric
\begin{equation}\label{ehlers}
{\rm d}s^2={\rm d}t^2-L^2e^{2\beta}{\rm d}x^2-L^2e^{-2\beta}{\rm
d}y^2-{\rm d}z^2,
\end{equation}
where $L$ and $\beta$ depend on $t-z$ or on $t+z$. The function $L$,
describing the transversal area, is called ``background factor", and
$\beta$, describing the deformation of a ring of test particles in
the transversal $(x,y)$ plane by gravitational forces, is called
``wave factor". The one and only Einstein equation\footnote{In
\cite{MTW} an abbreviated method to derive this Einstein equation
with the aid of a geodesic Lagrangian can be found (Box 14.4).} for
this system is
\begin{equation}\label{1stein}
L''+(\beta')^2L=0.
\end{equation}
It determines the background, when a wave factor is given, or vice
versa, so one function is left free, up to the condition that $L$
and $L''$ must have opposite signs. In consequence, a positive $L$
in front of a wave pulse leads necessarily to $L=0$ somewhere inside
or behind the pulse and the coordinate system becomes invalid there.
There is, however, no curvature singularity associated with it. For
a regular coordinate system behind the wave, see \cite{MTW}.
Gravitational waves of his kind, traveling exactly at the speed of
light along $z$, do not interact with each other and thus can be
superimposed like in a linear theory.

\subsection{The Gowdy Model}
The following treatise on plane gravitational waves is formally
based on a cosmological Gowdy model \cite{G}. Gowdy metrics are
exact solutions of the Einstein equations, they represent closed
universes with various topologies ($S^3$, $S^1\times S^2$, $T^3$),
containing gravitational waves along a single direction. There are
two Killing vectors transversal to the direction of propagation,
generating a commutative two-parameter group, which acts on the
spatial slices.

In our case we will assume a torus $T^3$ model to be opened up to a
cylindrical one and introduce coordinates $t$, $x$, $y$, $z$, with
the waves propagating in the $z$ direction, which is orthogonal to
$x$ and $y$. Accordingly, we introduce adapted densitized triad
variables
\begin{equation}\label{EEE}
{E^z}_3=:{\cal E},\hspace{1cm}
\begin{array}{rr} {E^x}_1=E^x\cos\beta, &{E^x}_2=E^x\sin\beta,\\
-{E^y}_1=E^y\sin\beta, &{E^y}_2=E^y\cos\beta,
\end{array}
\end{equation}
whereas $${E^z}_1={E^z}_2={E^x}_3={E^y}_3=0.$$ $E^x$ and  $E^y$ are
radial densitized triad components,
\begin{equation}
E^a=\sqrt{({E^a}_1)^2+({E^a}_2)^2}, \hspace{2cm}a=x,y,
\end{equation}
in the $x$ and $y$ direction, and $\cal E$ is the corresponding
quantity in the $z$ direction. When $x$ and $y$ are assumed
orthogonal, as in the present case, the model is a polarized one,
which is much easier to handle.

Gauge rotations around the $z$ axis are generated by the Gau\ss\
constraint
\begin{equation}
G=\partial_z{\cal E}+P^\beta,
\end{equation}
where $P^\beta$ is the conjugate momentum to the angle $\beta$ in
(\ref{EEE}). As the Gau\ss\ constraint and the corresponding
condition $\beta=0$ have vanishing Poisson brackets with the other
constraints, we can set them both strongly equal to zero, so that
$G$ will not be present any more. What remains are $\cal E$ and the
radial variables $E^x$ and $E^y$, and the corresponding connection
variables $\cal A$, $K_x$, $K_y$, (see (\ref{K}) below) subject to
the Hamiltonian and the diffeo constraint.

\subsection{Generalization, Ashtekar variables}
The metric components $g_{tt}$ and $g_{zz}$ in (\ref{ehlers}) being
equal to one, this model does not display any longitudinal
fluctuations. In a quantum description, however, we would expect
geometric fluctuations in every direction, even in flat space
without gravitational waves. For this reason we start from a
slightly more general classical point of view, taken from a
polarized Gowdy model and a quantum version thereof \cite{BD1,BD2}.
The metric for a polarized model in terms of the adapted densitized
triad variables introduced in (\ref{EEE}) is given by
\begin{equation}\label{metric}
{\rm d}s^2=-N^2{\rm d}t^2+{\cal E}\frac{E^y}{E^x}\,{\rm d}x^2+{\cal
E}\frac{E^x}{E^y}\,{\rm d}y^2+\frac{E^xE^y}{\cal E}\,{\rm d}z^2.
\end{equation}
All the triad components, as well as the lapse function $N$, are
assumed to be functions of $t$ and $z$ (not necessarily only of a
certain combination of $t$ and $z$ from the beginning). The model is
effectively 1+1-dimensional.

In the case $N^2=g_{zz}$ and when the metric functions depend only
on $t-z$, the Einstein equation corresponding to (\ref{1stein}) is
\begin{equation}\label{2stein}
\frac{{\cal E}''}{\cal E}+\frac{1}{2}\left(\frac{{\cal E}'}{\cal
E}\right)^2-(\ln E^xE^y)'\,\frac{{\cal E}'}{\cal
E}+\frac{1}{2}\left(\left(\ln\frac{E^y}{E^x}\right)'\right)^2=0.
\end{equation}
For comparison with (\ref{1stein}) we set $\sqrt{\cal E}=L$ and
$\;\ln(E^y/E^x)=2\beta$, then we find
\begin{equation}\label{3stein}
L''-(\ln E^xE^y)'\,L^2\,L'+(\beta')^2\,L=0,
\end{equation}
which reduces to (\ref{1stein}), when $E^xE^y=$ const. In
(\ref{3stein}) the second derivative of $L$ does not necessarily
have the opposite sign as $L$, in contrast to (\ref{1stein}).

The canonically conjugate connection to $E^x$, $E^y$, and $\cal E$
is derived from the (diagonal) extrinsic curvature
\begin{equation}\label{ext}
K_{aa}=\frac{1}{2N}\,\dot g_{aa}.
\end{equation}
The sign convention is the same as in \cite{BD2}. With the aid of
the triad covectors ${e^i}_b$ we construct the connection components
${K_a}^i={e^i}_b\,{K_a}^b$, in the notation of \cite{BD1,BD2}, and
find the radial components $K_a=\sqrt{({K_a}^1)^2+({K_a}^2)^2}\,$
with $a=x,y$, relevant for the polarized model, and $\cal A$,
\begin{equation}\label{K}
K_x=\frac{1}{N}\,\partial_t\sqrt{{\cal
E}\frac{E^y}{E^x}},\hspace{8mm}
K_y=\frac{1}{N}\,\partial_t\sqrt{{\cal
E}\frac{E^x}{E^y}},\hspace{8mm} {\cal
A}=\frac{1}{N}\,\partial_t\sqrt{\frac{E^xE^y}{\cal E}}
\end{equation}
with the nonzero Poisson brackets
\begin{equation}
\{K_a(z),E^b(z')\}=\kappa\,{\delta_a}^b\,\delta(z-z'),
\hspace{5mm}\{{\cal A}(z),{\cal E}(z')\}=\kappa\,\delta(z-z').
\end{equation}
$\kappa=2G_{\rm N}/\pi$, $G_{\rm N}$ is Newton's gravitational
constant. With the rotations around the $z$ axis being fixed by the
Gau\ss\ constraint at an early stage, we have only the canoni\-cal
variables introduced above, and the Hamiltonian constraint and one
diffeo constraint, generating diffeos along $z$:
\begin{eqnarray}\label{H}
H=-\frac{1}{\kappa}\frac{1}{\sqrt{{\cal
E}E^xE^y}}\left[K_xE^xK_yE^y+(K_xE^x+K_yE^y){\cal
AE}-\rule{0mm}{6mm}\right.\hspace{5mm}\label{constraint}\\
\;\;\;\;-\left.\frac{1}{4}\,({\cal E}')^2-\left(\frac{1}{2}{\cal
E}\left(\ln\frac{E^y}{E^x}\right)'\right)^2+\frac{1}{2}{\cal
EE}'(\ln
E^xE^y)'-{\cal EE}''\right]=0\nonumber,\\[3mm]
D=\frac{1}{\kappa}\,(K_x'E^x+K_y'E^y-{\cal AE}')=0.\hspace{45mm}
\end{eqnarray}

\subsection{One-way Killing conditions}
As it was stated in the Introduction, Gowdy models are characterized
by a two-parameter Abelian isometry group of transversal
translations in addition to symmetry of GR under spatial
diffeomorphisms and time evolution. These isometries are already
implemented in the form of the metric (\ref{metric}). Physically the
models describe plane waves propagating in one or in two opposite
directions, the latter case leading to a singularity. Space-times
with unidirectional waves, propagating at the speed of light, are
isometric under translations in a null direction along the waves. In
other words, for the $z$ coordinate along the direction of
propagation, there exists a null Killing vector field $k^\mu$ with
the corresponding covector
\begin{equation}\label{covector}
k_\mu=\left(Nk,0,0,\pm\sqrt{\frac{E^xE^y}{\cal E}}\,k\right),
\end{equation}
$k$ is a nonzero function  of $t$ and $z$. The plus/minus sign
determines plane waves into the positive/negative $z$ direction.
This Killing field commutes with the transversal Killing fields,
generating the above-mentioned two-parameter group, so the algebra
of these three Killing fields is commutative. The relation to the
symmetry constraints of GR will be given in form of Poisson brackets
in subsection 2.6.

The Killing equation $k_{0;0}=0$ leads to
\begin{equation}\label{kdot}
\dot k=\pm\sqrt{\frac{\cal E}{E^xE^y}}\,N'k,
\end{equation}
the equation $k_{3;3}=0$ to
\begin{equation}\label{kprime}
k'=\pm\frac{1}{N}\left(\partial_t\sqrt{\frac{E^xE^y}{\cal
E}}\right)k,
\end{equation}
where dot and prime mean derivatives w.\,r. to $t$ and $z$. Equating
the $z$ derivative of $(\ln k)\dot{}$\, from (\ref{kdot}) with the
$t$ derivative of $(\ln k)'$ from (\ref{kprime}) yields the
integrability condition
\begin{equation}\label{int}
\left(\frac{N'}{\sqrt{g_{zz}}}\right)'=\left(\frac{\dot{\sqrt{g_{zz}}}}{N}\right)\!\!\raisebox{4mm}{$\dot{}$}.
\end{equation}
This is a condition for the existence of a function $k$ such that
(\ref{covector}) is a Killing covector.

The equation $k_{0;3}+k_{3;0}=0$ does not give rise to further
relations, with (\ref{kdot}) and (\ref{kprime}) it is identically
fulfilled. Assuming some lapse function $N$ means to choose a time
unit, then (\ref{int}) fixes the dependence of $g_{zz}$ on the
coordinate time $t$ and on $z$. For the choice $N=\sqrt{g_{zz}}\;$
it gives
$$\left(\ln g_{zz}\right)''=\left(\ln g_{zz}\right)\ddot{}$$
with the solution
$$\ln g_{zz}=f(t+z)+g(t-z),$$
leading to the following form of the integrability condition
\begin{equation}\label{gzz}
g_{zz}=\frac{E^xE^y}{\cal E}=F(t+z)\cdot G(t-z).
\end{equation}
This condition is only necessary, not sufficient; for the actual
existence of a null Killing vector field in one direction, either
$F$ or $G$ must be constant.

The remaining two non-trivial Killing equations $k_{1;1}=0$ and
$k_{2;2}=0$ lead to the two conditions that determine unidirectional
plane waves. They are the subject of this and the next three
sections. With the time derivatives in the Christoffel symbols
$\Gamma^0_{11}$ and $\Gamma^0_{22}$ expressed in terms of $K_x$ and
$K_y$ with the help of (\ref{K}) they assume the forms:
\begin{equation}
\Gamma_{11}^0=\frac{1}{2N^2}\partial_t\left({\cal
E}\frac{E^y}{E^x}\right)=\frac{K_x}{N}\sqrt{{\cal E}\frac{E^y}{E^x}}
\end{equation}
and
\begin{equation}
\Gamma_{22}^0=\frac{1}{2N^2}\partial_t\left({\cal
E}\frac{E^x}{E^y}\right)= \frac{K_y}{N}\sqrt{{\cal
E}\frac{E^x}{E^y}}.
\end{equation}
When these expressions and (\ref{kdot}) and (\ref{kprime}) for the
derivative of $k$ are inserted into $k_{1;1}=0$ and $k_{2;2}=0$, $k$
drops out and this gives relations independent also of $N$,
\begin{eqnarray}
\label{Ux}U_x:=E^xK_x+\frac{1}{2}\,{\cal E}'+\frac{1}{2}\,{\cal
E}\left(\ln\frac{E^y}{E^x}\right)'=0,\label{x}\\
\label{Uy}U_y:=E^yK_y+\frac{1}{2}\,{\cal E}'-\frac{1}{2}\,{\cal
E}\left(\ln\frac{E^y}{E^x}\right)'=0.\label{y}
\end{eqnarray}
With this choice of signs, these are one-way conditions for waves
propagating into the positive $z$ direction
($\partial_t=-\partial_z$) in terms of the fundamental variables of
the system. (\ref{x}) and (\ref{y}) are conditions on the variables
of the system for the existence of a null Killing field, they do not
specify a concrete form of the function $k$.

As already mentioned, the integrability condition (\ref{gzz}) does
not uniquely determine left- or right-going waves, they must be
picked out by hand. In view of this, one might also choose a more
elementary, alternative approach: Postulating that the derivative of
the metric components $g_{xx}$ and $g_{yy}$ w.\,r. to the physical
time $Nt$ is equal to minus the derivative w.\,r. to the physical
length $\sqrt{g_{zz}}\,z$,
\begin{equation}
\sqrt{\frac{E^xE^y}{\cal E}}\,\dot g_{aa}=-Ng_{aa}',
\hspace{10mm}a=1,2
\end{equation}
leads to (\ref{Ux}) and (\ref{Uy}), when the time derivatives are
expressed in terms of $K_x$ and $K_y$. When $N$ is chosen to be
equal to $\sqrt{E^xE^y/{\cal E}}$, for a wave traveling into the
positive $z$ direction, this gives the further relation
\begin{equation}
\label{Uz} U_z:={\cal AE}-\frac{1}{2}\,{\cal E}'+\frac{1}{2}\,{\cal
E}\left(\ln E^yE^x\right)'=0
\end{equation}
as a consequence of the integrability condition (\ref{int}), but as
this is dependent on a choice of the lapse function, it is not an
independent Killing condition and not necessary in the following. In
the present approach $U_z$ may be used later on just to make the
relations (\ref{HD}) and (\ref{hd}), appearing below, more
convenient. Namely, when we combine $U_x$, $U_y$, their derivatives
w.\,r. to $z$, and $U_z$, with the Hamiltonian and the diffeo
constraint, we obtain the following relation for waves in the
positive $z$ direction
\begin{equation}\label{HD}
\kappa\left(\sqrt{\frac{E^xE^y}{\cal E}}\,H+D\right)=-\frac{1}{\cal
E}\left(U_xU_y+U_xU_z+U_yU_z\right)+(U_x+U_y)'.
\end{equation}
We see that the r.\,h. side is equal to zero, when $U_x$ and $U_y$
are zero, independently of $U_z$. The expression (\ref{HD}) has
density weight two. $\sqrt{E^xE^y/{\cal E}}\:H$ is the evolution
generator in the case of the special lapse function
$N=\sqrt{g_{zz}}$. $\;U_x\equiv0$ and $U_y\equiv0$ are sufficient
for the essential equivalence of the Hamiltonian with the diffeo
constraint, $\sqrt{E^xE^y/{\cal E}}\:H=\pm D$, i.\,e. the strict
proportionality of the time and the $z$ derivatives, whereas $U_z=0$
is not needed, it just makes the relation look nicer.

\subsection{Einstein equations, massless fields}

When the metric components in (\ref{metric}) are merely functions of
$t$ and $z$ and not only of $t\pm z$, but the lapse function is
chosen to be $N=\sqrt{\frac{E^xE^y}{\cal E}}$, the Einstein tensor
has five independent non-vanishing components, the diagonal ones and
$G_{03}=G_{30}$, giving rise to five equations for the six triad and
connection variables. $G_{00}$ and $G_{33}$ do not contain time
derivatives of the phase space variables, so in the case of vacuum
they represent the Hamiltonian and the diffeo constraint,
\begin{equation}
G_{00}=-\frac{1}{\cal E}\sqrt{\frac{E^xE^y}{\cal E}}\,H,
\hspace{1cm} G_{03}=-\frac{1}{\cal E}\,D.
\end{equation}
In the presence of matter they represent the gravitational energy
and momentum density in the chosen reference frame. The remaining
Einstein equations are dynamical, they determine $\dot K_x$, $\dot
K_y$, and $\dot{\cal A}$ in terms of the fundamental geometrical
(and perhaps matter) variables, thus leading to a system of three
coupled nonlinear first-order PDEs for time evolution.

When, under the assumption $N=\sqrt{g_{zz}}$, the one-way conditions
(\ref{x},\ref{y},\ref{HD}) are satisfied, then all metric components
depend on $t-z$, and the Einstein tensor reduces to three identical
nonzero components,
\begin{equation}\label{gab}
G_{00}=G_{33}=-G_{03},
\end{equation}
(see (\ref{2stein})) of constraint type, no further nontrivial
components arise. - With the above assumptions the time evolution is
already fixed.  If, in addition, the vacuum Einstein equation is
imposed on (\ref{gab}), one obtains pure unidirectional
gravitational waves and $G_{00}$ becomes the Hamiltonian constraint,
or the equivalent diffeo constraint, respectively.

In the presence of matter $G_{ab}$ is coupled to the energy-momentum
tensor of a field. As the trace of the Einstein tensor in
(\ref{gab}) is zero, with or without matter, the field must be
massless for consistency. Without imposing the vacuum constraint
$G_{00}=0$, the condition ``time derivative $=\pm z$ derivative"
does not distinguish between pure gravitational waves and the
gravitational
background accompanying massless fields. 

\subsection{Alternative set of conditions, physical meaning}
The sum and the difference of the two $N$-independent conditions
$U_x$ and $U_y$ for wave propagation into the positive $z$ direction
are
\begin{eqnarray}
&&U_+:=K_xE^x+K_yE^y+{\cal E}'=0,\label{Up}\\
&&U_-:=K_xE^x-K_yE^y+{\cal
E}\left(\ln\frac{E^y}{E^x}\right)'=0.\label{Um}
\end{eqnarray}
With $K_x$ and $K_y$ from (\ref{K}) inserted, $U_+=0$ means
\begin{equation}
\sqrt{\frac{E^xE^y}{\cal E}}\,\dot{\cal E}=-N{\cal E}',
\end{equation}
namely the physical time derivative of $\cal E$ is equal to the true
physical spatial derivative along $z$. This refers to the background
factor, the transversal area $\cal E$. In the same way $U_-=0$ means
\begin{equation}
\sqrt{\frac{E^xE^y}{\cal
E}}\,\left(\ln\frac{E^y}{E^x}\right)\!\!\raisebox{3mm}{$\dot{}$}=-
N\left(\ln\frac{E^y}{E^x}\right)',
\end{equation}
indicating equivalence of time and space derivatives of the wave
factor $\ln(E^y/E^x)$.

As (\ref{HD}) can also be written in the form
\begin{equation}\label{hd}
\kappa\left(\sqrt{\frac{E^xE^y}{\cal E}}\,H+ D\right)=\frac{1}{4\cal
E}\,U_-^2-\frac{1}{4\cal E}\,U_+^2-\frac{1}{\cal E}\,U_+U_z+U_+',
\end{equation}
it follows from the equivalence of $H$ and $D$ (particularly when
$H=0$ and $D=0$) that $U_-=0$, whenever $U_+=0$, thus the wave
factor propagates at the speed of light, when the background factor
does. As linear combinations of each other, the pairs $(U_x,U_y)$
and $(U_+,U_-)$ are, of course, equivalent sets of conditions. In
the following we will mainly work with $U_\pm$.

The one-way conditions Poisson-commute weakly with the constraints
of General Relativity (see next subsection), $U_+$ also with $U_-$
and with itself, $U_-$ does not commute with itself in general. In
detail the Poisson brackets between the one-way conditions are the
following:
\begin{eqnarray}
&&\{U_x[f],U_x[g]\}=\{U_y[f],U_y[g]\}=\frac{1}{2}\,{\cal
E}[f'g-fg'],\nonumber\\
&&\{U_x[f],U_y[g]\}=\frac{1}{2}\,{\cal E}[fg'-f'g],\\
&&\{U_+[f],U_+[g]\}=\{U_+[f],U_-[g]\}=0,\nonumber\\
&&\{U_-[f],U_-[g]\}=2\,{\cal E}[f'g-fg'],\nonumber
\end{eqnarray}

As can be seen, some of the conditions do Poisson-commute only for
proportional test functions $f$ and $g$. As one-way conditions are
not gauge constraints, we are only looking for a certain class of
solutions, not for a theory with additional symmetries, so the
Poisson brackets between the conditions do not have to be imposed as
further conditions. In quantum theory the nonzero commutators will
not be relevant as operators, in other words, we do not have to
enlarge the algebra of constraints by the one-way conditions, this
would over-constrain the system.

\subsection{Relation to the constraints of GR}
$U_+$ and $U_-$, and even $U_z$, as densities of weight one, do
weakly Poisson-commute with the diffeo constraint,
\begin{equation}
\{U_\pm[f],D[g]\}=-U_\pm[f'g],\hspace{5mm}\{U_z[f],D[g]\}=-U_z[f'g].
\end{equation}
The Poisson brackets with the Hamiltonian constraint are the
following
\begin{equation}
\{U_+[f],H[g]\}=U_+\left[\sqrt{\frac{\cal
E}{E^xE^y}}\,f'g\right]-H[fg]=\sqrt{\frac{\cal
E}{E^xE^y}}\:U_+[f'g]-H[fg],
\end{equation}
\begin{equation}
\{U_-[f],H[g]\}=\frac{1}{\sqrt{{\cal E}E^xE^y}}\left(U_-\left[({\cal
E}f)'g\right]-{\cal
E}\left(\ln\frac{E^y}{E^x}\right)'U_+[fg]\right).
\end{equation}
$U_+$ weakly commutes with $H$, the condition $U_-$ for the wave
factor needs also the condition $U_+=0$ for the background factor to
be satisfied, in order to commute with $H$. Just for completeness:
the coordinate dependent condition $U_z$ does not commute with $H$
with arbitrary lapse functions, it does with $N=\sqrt{g_{zz}}$,
\begin{eqnarray}
&&\left\{U_z[f],H\left[\sqrt{\frac{E^xE^y}{\cal
E}}\right]\right\}=-U_z\,[f']+\frac{1}{\cal
E}\,(U_xU_y+U_xU_z+U_yU_z)\,[f]-\\
&&-\left(\frac{1}{2}\,(\ln g_{xx})'\,U_x+\frac{1}{2}\,(\ln
g_{yy})'\,U_y+2\,(\ln{\cal E})'\,U_z\right)[f].\nonumber
\end{eqnarray}
Summarizing the classical part, we state that there are two Killing
conditions for making time evolution equivalent to space
translation. In the following sections we shall investigate in how
far these relations carry over to quantum theory.

\section{Quantum states and operators}
Quantum states of the present model are one-dimensional spin network
(SNW) functions on a sequence of edges and vertices along the $z$
axis, made from point holonomies at vertices and the usual path
holonomies along edges. At a vertex $v$ the former ones are
\begin{equation}\label{38}
|m\rangle=e^{i\frac{m}{2}K_x(v)} \hspace{5mm}\mbox{and}\hspace{5mm}
|n\rangle=e^{i\frac{n}{2}K_y(v)}
\end{equation}
with integer $m$ and $n$. 
In the present one-dimensional SNWs vertex functions in form of (a
pair of) point holonomies replace loop holonomies of $K_x$ and $K_y$
in the homogeneous $x$ and $y$ direction.

In \cite{BD2} the state functions are elements of the Bohr
compactification of the reals (\cite{Bohr}, Chapter 28), with
representation labels $m$ and $n$. But in fact all these holonomies
are phases and in the following Sections 4 - 6 we will need only
sequences of integers $m$ and $n$, determined by the raising and
lowering parts of the connection operators, see (\ref{conoperator})
below. So nothing is lost when we dispense with ${\bf R}_{\rm Bohr}$
and reduce our interest until further notes effectively to $U(1)$
representations with the labels $m/2$ and $n/2$.

Edge holonomues are
\begin{equation} |k\rangle=e^{i\frac{k}{2}\int_e{\cal A}\,{\rm d}z}
\end{equation}
where $e$ denotes an edge and $k$ is an integer number, also
denoting a representation of $U(1)$ by its very nature. $k$ has the
meaning of a quantum number of transversal area, as well as $m$ and
$n$ are area quantum numbers in the $(y,z)$, and the $(x,z)$ plane,
respectively.

A basic state for the whole system is a tensor product over all
edges and vertices of two point holonomies for each vertex $v_i$ and
an edge holonomy for every edge $e_j$:
\begin{equation}|\{m_i,n_i\},\{k_j\}\rangle=
\prod_ie^{i\frac{m_i}{2}K_x(v_i)}\,e^{i\frac{n_i}{2}K_y(v_i)}\prod_je^{i\frac{k_j}{2}\oint_{e_j}{\cal
A}\,{\rm d}z}.
\end{equation}
$\{m_i,n_i\}$ and $\{k_j\}$ denote holonomy label sets along the
whole SNW, characterizing a state.

The connection components are promoted to holonomy operators. As in
the present model all holonomies are mere phase factors, we will, in
contrast to \cite{BD2}, where holonomy operators act as traces of
$SU(2)$ operators on states, introduce $U(1)$ vertex operators by
first setting
\begin{equation}\label{conoperator}
K_a\rightarrow\hat{K}_a=2\sin\left(\frac{1}{2}K_a
\right)=\frac{1}{i}\left(e^{\frac{i}{2}K_a}-e^{-\frac{i}{2}K_a}\right).
\end{equation}
($K_a$ without argument $v_i$ is the general form at any vertex.)
When $\hat{K}_x$ acts on a holonomy $|m\rangle$, this gives
\begin{equation}
\frac{1}{i}(|m+1\rangle-|m-1\rangle),
\end{equation}
so $\hat{K}_x$ acts simply as a raising and lowering operator. Also
here we content ourselves with a lower mathematical level in favor
of clearer and simpler calculations. (For comparison: in \cite{we}
calculations like the following ones in Section 4 were done with
traces of $SU(2)$ operators, leading to qualitatively the same
outcomes.) The same is carried out with $K_y$, so that
\begin{equation}
\hat{K}_y|n\rangle=\frac{1}{i}(|n+1\rangle-|n-1\rangle),
\end{equation}
if $|n\rangle$ is a holonomy of $K_y$.

$E^x$ and $E^y$ (in principle functional derivatives $\delta/\delta
K_a$) are formulated as derivatives $\partial/\partial K_a$, so that
the densitized triads act as multiplication operators, when
integrated over some interval $\cal I$ containing a vertex
\begin{equation}
\int_{\cal I}{\rm d}z\,\hat E^x(z)|m\rangle=\frac{\gamma l_{\rm
P}^2}{2}\,m|m\rangle, \hspace{10mm} \int_{\cal I}{\rm d}z\,\hat
E^y(z)|n\rangle=\frac{\gamma l_{\rm P}^2}{2}\,n|n\rangle.
\end{equation}
$l_{\rm P}$ is the Planck length, $\gamma$ is the Immirzi parameter,
$\gamma l_{\rm P}^2$ is the typical scale of discreteness of both
$|m\rangle$ and $|n\rangle$.

Analogously $\cal A$ is promoted to an operator; as it is a scalar
density we have
\begin{equation}
{\cal A} \rightarrow 2\sin\left(\frac{1}{2}\int_e{\cal A}\,{\rm
d}z\right)=\frac{1}{i}\left(e^{\frac{i}{2}\int_e{\cal A}\,{\rm
d}z}-e^{-\frac{i}{2}i\int_e{\cal A}\,{\rm d}z}\right)
\end{equation}
with its action on edge holonomies
\begin{equation}
\hat{\cal
A}\,|k\rangle=\frac{1}{i}\left(|k+1\rangle-|k-1\rangle\right),
\end{equation}
and the conjugate multiplication operator $\cal E$ acts as
\begin{equation}\label{calE}
\hat{\cal E}|k\rangle=\frac{\gamma l_{\rm P}^2}{2}\,k\,|k\rangle.
\end{equation}
So, $\hat{\cal E}$ is essentially the transversal area operator in
units of $\gamma l_{\rm P}^2/2$, $\hat{V}=\widehat{\sqrt{{\cal
E}E^xE^y}}$ is the volume operator. In the following $\gamma l_{\rm
P}^2/2$ will be set equal to one.

\section{The one-way conditions as operators on quantum state
functions, solutions}

\subsection{An operator corresponding to $U_+$}

The expressions $K_xE^x$ and $K_yE^y$, containing connection
components, act nontrivially on vertex holonomies, but leave edge
holonomies unchanged, so we can assume the latter ones as
arbitrarily given. $U_+$, commuting with itself, is easier to
promote to an operator than $U_-$, it can naturally be associated to
each vertex. First, we choose a symmetric factor ordering,
\begin{equation}\label{+}
U_+=\sqrt{E^x}\,K_x\,\sqrt{E^x}+\sqrt{E^y}\,K_y\,\sqrt{E^y}+{\cal
E}'.
\end{equation}
The derivative term ${\cal E}'$ is naturally interpreted as
$(k_+-k_-)/\epsilon$, where $k_\pm$ are the edge labels right and
left from the considered vertex, and $\epsilon$ is the length of the
same interval $\cal I$ as in the definition of the triad operators.
To obtain an operator from the density expressions in (\ref{+}), the
equation is integrated over $\cal I$, so ${\cal E}'$ gives simply
the difference $k_+-k_-$. Taking a (double) vertex holonomy
\begin{equation}\label{amn}
|m,n\rangle:=|m\rangle\otimes|n\rangle,
\end{equation}
the action of $\hat{U}_+:=\int_{\cal I}{\rm d}z\,U_+(z)$, with
$\gamma l_{\rm P}^2=1$, on such a holonomy is the following
\begin{eqnarray}\label{49}
&&\hat{U}_+|m,n\rangle=\frac{1}{i}\left[\sqrt{m(m+1)}\,|m+1,n\rangle-\sqrt{m(m-1)}\,
|m-1,n\rangle\right.\\
&&\hspace{10mm}+\left.\sqrt{n(n+1)}\,|m,n+1\rangle-\sqrt{n(n-1)}\,|m,n-1\rangle\right]+
(k_+-k_-)\,|m,n\rangle.\nonumber
\end{eqnarray}

\subsubsection{The difference equation and solutions}

For a vertex state of the form
\begin{equation}\label{psi}
|\psi\rangle=\sum_{m,n}a_{m,n}\,|m,n\rangle
\end{equation}
the condition $\hat{U}_+\,|\psi\rangle=0$ leads to the difference
equation
\begin{eqnarray}\label{DE}
&&\sqrt{m(m+1)}\,a_{m+1,n}-\sqrt{m(m-1)}\,a_{m-1,n}+\sqrt{n(n+1)}\,a_{m,n+1}-\sqrt{n(n-1)}\,a_{m,n-1}\nonumber\\[2mm]
&&=i\,b\,a_{m,n}.
\end{eqnarray}
$b$, coming from the difference of the cross section area (the
background), left and right, is defined as
\begin{equation}
b:=(k_+-k_-).
\end{equation}

As an introduction, in the following we consider in detail the
simplest case with $b=0$, indicating the same cross-section area to
the left and to the right of the given vertex (in classical GR
exactly constant cross section would mean no waves at all).

The partial difference equation (\ref{DE}) is separable and has a
one-parameter family of solutions. When separated, the parameter is
the separation constant. Here we prefer not to separate, but to
parametrize the solutions by initial values $a_{m,m}$ along the
diagonal and the differences $a_{m,m-1}-a_{m-1,m}$ of coefficients
above and below the diagonal in an $(m,n)$ diagram. The advantage is
some systematics: initial values on the diagonal give rise to
symmetric solutions, the differences lead to antisymmetric ones.

In the following we give two different symmetric and one
antisymmetric explicit solution. Inserting the values $m=0$ and
$n=0$ into (\ref{DE}) with $b=0$ gives identically zero, so we can
choose the coefficients $a_{1,0}$ and $a_{0,1}$ (and also $a_{0,0}$)
arbitrarily, setting them equal to zero, we get $a_{0,2}=0$ from
(\ref{DE}) with $m=0$, $n=1$, and so on. (With $b\neq0$ eq.
(\ref{DE}) fixes $a_{0,0}=0$.) In this way all $a_{m,0}$ and
$a_{0,n}$ become zero and we can leave zero-volume states aside and
restrict our attention to the quadrant ($m\geq1$, $n\geq1$) in a
discrete coordinate system of pairs $(m,n)$.

Setting $(m,n)=(1,1)$ we find $a_{1,2}+a_{2,1}=0$, from $(2,2)$
follows $a_{2,3}+a_{3,2}=0$, and so on. This means that we can
conveniently choose the diagonal elements $a_{m,m}$ as free initial
values, with the symmetric parts $a_{m+1,m}+a_{m,m+1}$ immediately
beside the diagonal determined to be zero by the difference
equation, whereas the antisymmetric parts remain undetermined. So,
from initial values $a_{m,m}$ with $a_{m,m\pm1}=0$ we can construct
symmetric solutions, on the other hand, from $a_{m,m}=0$ and
arbitrary antisymmetric initial values $a_{m,m-1}=-a_{m-1,m}$ we get
antisymmetric solutions.

\subsubsection{Three explicit solutions for $b=0$}

Let's consider the simplest case of a symmetric solution coming from
the initial value $a_{1,1}=1$, with all other diagonal values and
the values $a_{m,m\pm1}$ above and below the diagonal being zero.
Inserting $(2,1)$ into (\ref{DE}) gives
\begin{equation}
a_{3,1}=\frac{1}{\sqrt{3}}=\frac{2\sqrt{1\cdot3}}{1\cdot2\cdot3},
\end{equation}
for $(3,2)$ we get
\begin{equation}
a_{4,2}=\frac{2\sqrt{2\cdot4}}{2\cdot3\cdot4}.
\end{equation}
From these coefficients and a few further ones one can easily read
off the general formula
\begin{equation}
a_{n+2,n}=\frac{2\sqrt{n(n+2)}}{n(n+1)(n+2)}.
\end{equation}
Inserting $(n+2,n)$ gives $a_{n+3,n}=0$ $\forall n$, for $(n+3,n)$
inserted into (\ref{DE}) we obtain
\begin{equation}
a_{n+4,n}=\frac{4\sqrt{n(n+4)}}{(n+1)(n+2)(n+3)},
\end{equation}
for $(n+5,n)$ we find
\begin{equation}
a_{n+6,n}=\frac{6\,\sqrt{n(n+6)}}{(n+2)(n+3)(n+4)},
\end{equation}
from which we may read off the general form for $m-n$ even
\begin{equation}\label{58}
a_{m,n}=\frac{8\,(m-n)\,\sqrt{mn}}{(m+n-2)(m+n)(m+n+2)},
\end{equation}
where $n$ was set equal to $\frac{m+n}{2}-\frac{m-n}{2}$ in the
denominator, and prove by induction. With the definitions
$p=\frac{m+n}{2}$ and $q=\frac{m-n}{2}$, this becomes
\begin{equation}\label{pq}
a_{p,q}=\frac{2q\,\sqrt{p^2-q^2}}{(p-1)p(p+1)}.
\end{equation}

In the same way from the only nonzero initial value $a_{2,2}=1$ one
obtains the slightly more complicated symmetric solution
\begin{equation}\label{60}
a_{p,q}=\frac{4q\,(p^2-3q^2-1)\,\sqrt{p^2-q^2}}{p\,(p^2-1)\,(p^2-4)}.
\end{equation}
For large $p$ both these solutions go as $p^{-2}$. In the symmetric
solutions for all nonzero coefficients $a_{m,n}$ the sum $m+n$ is
even, in the antisymmetric solutions only coefficients with odd sum
$m+n$ are nonzero, in both cases there is a chessboard like pattern
in the discrete $(m,n)$ diagram.

As an antisymmetric example we present the solution coming from
$a_{2,1}=-a_{1,2}=1$ with all other possible initial values being
zero. For $m>n$, $m-n$ odd, it is
\begin{equation}
a_{m,n}=\frac{6\sqrt{2}[(m-n)^2-1]\,\sqrt{mn}}{[(m+n)^2-1][(m+n)^2-9]},
\end{equation}
or
\begin{equation}\label{21}
a_{p,q}=\frac{6\sqrt{2}\,(4q^2-1)\,\sqrt{p^2-q^2}}{(4p^2-1)(4p^2-9)}
\end{equation}
(for half-integer $q$), respectively. Here the coefficients go as
$p^{-3}$ for large $p$. The nontrivial case with $b\neq0$, where
symmetric and antisymmetric solutions get inevitably mixed, will be
treated together with $U_-$. Before doing this we have a look at the
normalizability problem of the solutions.

\subsubsection{Normalizability}
With the obvious inner product $\langle
m,n|m',n'\rangle=\delta_{m,m'}\,\delta_{n,n'}$ of holonomies, which
offers itself at first sight, the norm of a solution $|\psi\rangle$
(\ref{psi}) is equal to the sum of $a_{m,n}^2$ over $m>0$ and $n>0$.
Relabeled with $p$ and $q$ and reordered, the norm square of
solution (\ref{58}), or in the form (\ref{pq}), respectively, is
\begin{equation}\label{normsquare}
\langle\psi|\psi\rangle=1+\sum_{p=2}^\infty\,2\sum_{q=1}^{p-1}a_{p+q,p-q}^2.
\end{equation}
The finite sum over $q$ for fixed $p$ is
\begin{equation}\label{sum}
\sum_{q=1}^{p-1}\frac{4q^2(p^2-q^2)}{p^2(p^2-1)^2}=\frac{4}{(p^2-1)^2}\sum_{q=1}^{p-1}q^2-
\frac{4}{p^2(p^2-1)^2}\sum_{q=1}^{p-1}q^4.
\end{equation}
As the leading term of $\sum q^2$ is $(p-1)^3/3$, and that of $\sum
q^4$ is $(p-1)^5/5$, for large $p$ the sum (\ref{sum}) goes as
$\frac{8}{15}\,p^{-1}$, thus the norm square (\ref{normsquare})
diverges logarithmically. The same occurs with the solutions
(\ref{60}) and (\ref{21}). So, with the usual inner product in
$U(1)$ and the assumption of equal spacing of the eigenvalues of
geometric quantities the solutions are not normalizable. Decreasing
spacing for growing eigenvalues, as introduced, for example in
\cite{AP,Sak}, as lattice refinement, makes the situation worse.

Non-normalizable states, not lying in the Hilbert space of a system,
like the eigenstates of position or momentum, are not uncommon in
quantum mechanics. So it might appear that the above states, derived
from just one or two nonzero initial values, are examples of a
similar kind. The usual solution to such normalizability issues are
continuous wave packets. Here, however, for example the parameters
$a_{1,1}$, $a_{2,2}$, \ldots, characterizing symmetric solutions,
are discrete, as well as the parameters of the antisymmetric ones.
And discrete superpositions of functions of the form (\ref{38}) of
integers $m$ and $n$ are Fourier series of periodic functions, and
so the method of wave packets fails.

To understand the meaning of these states, recall that with zero
spatial derivative ${\cal E}'$ (in the discrete case with zero
difference $b=0$), $\hat{U}_+\,|\psi\rangle=0$ determines states
with local zero space and time derivatives of the background. This
satisfies one condition for quantum Minkowski space, but without
$\hat{U}_-\,|\psi\rangle=0$ it is not restrictive enough to give at
least an approximative model of flat space. On the other hand, with
zero difference of cross section areas, it deals only with
longitudinal fluctuations, as can be expected for the present
essentially $1+1$ - dimensional model.

The non-normalizability (under the assumption of the inner product
introduced above) points out a problem already at this early stage
and indicates that also more realistic solutions must be
renormalized in one or another way. This can be achieved by some
kind of cutoff in $p$ (suppression of large volumes/lengths) or in
$q$ (suppression of large wave factors), or in both directions. As
can be seen from (\ref{pq}, \ref{60}, \ref{21}), the coefficients
fall off sufficiently rapidly in the $p$-direction for fixed $q$,
whereas the sums over $q$ from 1 to $p-1$ are ``too large". By any
modification of the solution the condition
$\hat{U}_+\,|\psi\rangle=0$ becomes violated, so that the time
derivative of the background picks up fluctuations. Nevertheless,
for a cutoff in $p$ or a symmetric cutoff in $q$, at least the
expectation value of $\hat{U}_+$ remains zero, there are just
fluctuations in time.

\subsection{The operator $\hat{U}_-$}

\subsubsection{Difference equation for $\hat{U}_-$}
$U_-$ contains the derivative of $\ln(E^y/E^x)$, so it cannot be
associated as straightforwardly as $U_+$ to one and the same vertex,
the discrete version must involve at least one further vertex. Lets
denote the labels at the considered vertex  as $m_i$ and $n_i$, and
those left and right as $m_{i\pm1}$, $n_{i\pm1}$.

In the classical theory, when background and wave factor are
differentiable functions, it does not make a difference whether we
take the left or the right derivative. In the discrete case the
derivative of the wave factor can be interpreted in different
ways:\\

\noindent(i) Central difference
\begin{equation}\label{central}
\left(\ln\frac{E^y}{E^x}\right)'\rightarrow\ln\frac{n_{i+1}}{m_{i+1}}-\ln\frac{n_{i-1}}{m_{i-1}}.
\end{equation}
This seems to be the most natural interpretation.\\

\noindent(ii) Left and right difference, involving the local
variables $m_i$ and $n_i$
\begin{equation}\label{l+r}
\left(\ln\frac{E^y}{E^x}\right)'\rightarrow\ln\frac{n_i}{m_i}-\ln\frac{n_{i-1}}{m_{i-1}},\hspace{10mm}\mbox{or}\hspace{10mm}
\rightarrow\ln\frac{n_{i+1}}{m_{i+1}}-\ln\frac{n_i}{m_i}.
\end{equation}\\

For the choice (i) at each vertex the appertaining variables $m_i$
and $n_i$ do not occur in the difference equations, $\hat{U}_+$ and
$\hat{U}_-$ commute at each vertex and have a common solution for
each pair of neighboring vertex holonomies and each pair $(k_-,k_+)$
of adjacent edge labels. But for the construction of a global
solution at (some extended part of) the whole SNW, choice (i) is
quite inconvenient because of non-commutativity at neighboring
vertices, whereas for (ii) one could go from vertex to vertex, but
here it turns out that there is no common local solution. The
non-commutativity of the Poisson brackets of the classical $U_-$
with itself is thus reflected in the non-commutativity of the
corresponding operators at one and the same or at neighboring
vertices. Even when we first ignore global issues, choose version
(i), and calculate local solutions, normalization problems will
prevent us from accepting them as physical states.

For the expression $\cal E$ at a vertex we take the mean value of
the edge labels left and right:
\begin{equation}\label{Ek}
{\cal E}\rightarrow\frac{1}{2}\,(k_-+k_+)=:k
\end{equation}
for (i), for (ii) simply ${\cal E}\rightarrow k_-$ or ${\cal
E}\rightarrow k_+$ is convenient.\\

\noindent {\bf The difference equation}: For the choice (i) for
$\left(\ln(E^y/E^x)\right)'$ and (\ref{Ek}) for $\cal E$ the
difference equation derived from $U_-$ is the following
\begin{eqnarray}\label{U-}
&&\sqrt{m(m+1)}\,a_{m+1,n}-\sqrt{m(m-1)}\,a_{m-1,n}-\sqrt{n(n+1)}\,a_{m,n+1}+\sqrt{n(n-1)}\,a_{m,n-1}
\nonumber\\[2mm]
&&=iwa_{m,n},
\end{eqnarray}
where $m\equiv m_i$, $n\equiv n_i$, and
\begin{equation}\label{w}
w:=\frac{1}{2}(k_++k_-)\left(\ln\frac{n_{i+1}}{m_{i+1}}-\ln\frac{n_{i-1}}{m_{i-1}}\right)
\end{equation}
stands for (the derivative of the) wave factor and is constant at
the vertex under consideration, it depends only on neighboring
vertex and edge labels. (For the choice (ii) we would have to write
$w(m,n)$ instead of simply $w$.)

\subsubsection{Common solutions}
With the decision for $w$ in the form (\ref{w}), $\hat{U}_+$ and
$\hat{U}_-$ at each vertex are commuting Hermitian operators, so
there is a nontrivial one-parameter common solution. Operators at
neighboring vertices do not commute, but for the moment we
concentrate at a local solution at one vertex and ignore issues with
non-commutativity to come back to them later. The partial difference
equations (\ref{DE}) and (\ref{U-}) are in this case separable with
the product ansatz $a_{m,n}=\alpha_m\beta_n$, resulting in
\begin{equation}\label{UxUy}
\hat{U_x}\,\sum_m\alpha_m|m\rangle=0
\hspace{1cm}\mbox{and}\hspace{1cm}\hat{U_y}\sum_n\beta_n|n\rangle=0
\end{equation}
in terms of operators corresponding to the original expressions
$U_x$ and $U_y$, as $U_\pm=U_x\pm U_y$. The separated difference
equations are
\begin{eqnarray}
&&\sqrt{m(m+1)}\:\alpha_{m+1}-\sqrt{m(m-1)}\:\alpha_{m-1}-\frac{i}{2}\,(b+w)\,\alpha_m=0,\label{73}\\[3mm]
&&\label{74}\sqrt{n(n+1)}\:\beta_{n+1}-\sqrt{n(n-1)}\:\beta_{n-1}-\frac{i}{2}\,(b-w)\beta_n=0.
\end{eqnarray}

With $\alpha_1=\beta_1=1$ the first few coefficients in a solution
of (\ref{73}) are
\begin{eqnarray}\label{common}
&&\alpha_2=i\,\frac{b+w}{2\sqrt{2}},
\hspace{1cm}\displaystyle\alpha_3=\frac{1}{\sqrt{3}}\,\left(1-\frac{1}{8}\,(b+w)^2\right),\hspace{1cm}
a_4=i\frac{b+w}{6}\left(2-\frac{1}{8}\,(b+w)^2\right),\label{00}\nonumber\\[5mm]
&&\alpha_5=\frac{1}{\sqrt{5}}\left(1-\frac{5}{24}\,(b+w)^2+\frac{1}{192}\,(b+w)^4\right),\ldots.
\end{eqnarray}
and $\beta_n$ is of the same form as $\alpha_n$ with $b+w$ replaced
by $b-w$. The resulting coefficients $a_{m,n}=\alpha_m\,\beta_n$ are
symmetric in $b$ and antisymmetric in $w$.

To see the asymptotic behavior of the solution of (\ref{73}) and
(\ref{74}) we replace $\alpha_m$ by a differentiable function
$\alpha(m)$ and approximate the difference equation by a
differential equation:
\begin{equation}\label{alpha}
\sqrt{m}\;2\frac{\rm d}{{\rm
d}m}\left(\sqrt{m}\,\alpha(m)\right)=\frac{i}{2}(b+w)\,\alpha(m)
\end{equation}
with the solution
\begin{equation}\label{am}
\alpha(m)=\frac{1}{\sqrt{m}}\,\exp\left[\frac{i}{4}(b+w)\,\ln
m\right].
\end{equation}
Doing the same with the second equation gives
\begin{equation}\label{bn}
\beta(n)=\frac{1}{\sqrt{n}}\,\exp\left[\frac{i}{4}(b-w)\,\ln
n\right],
\end{equation}
so that the asymptotic behavior is
\begin{equation}\label{asymp}
a_{m,n}\rightarrow\frac{1}{\sqrt{mn}}\,\exp\left[\frac{i}{4}\left(b\,\ln(mn)+w\ln\frac{m}{n}\right)\right].
\end{equation}
As expected, this is not normalizable in the $U(1)$ inner product.

\section{Renormalization, expectation values}
\subsection{Norm}
The common solution with arbitrary $b$ and $w$, presented in the
foregoing section, contains the function $1/\sqrt{mn}$ plus powers
of $b\pm w$. First we will consider the simpler state with $b=w=0$
and $a_{m,n}=1/\sqrt{mn}$, introduced below as contribution to the
vacuum state (\ref{vak}), and then show that in the general case the
normalization problem is the same.

So far we had assumed the obvious inner product $\langle
m,n|m',n'\rangle=\delta_{m,m'}\delta_{n,n'}$ of holonomies (= phase
factors) for vertex states. However, basic states in full LQG are
$SU(2)$ holonomies, in the present context they are artificially
squeezed into $U(1)$ phase factors by symmetry reduction.

We might prefer an inner product of the basic holonomies close to
that of full LQG: There the basic SNW functions are the $SU(2)$
matrix elements ${D^j(g)^\mu}_\nu$ of a group element $g$ in a
representation labeled by some half-integer $j$. According to the
Peter-Weyl theorem \cite{PW} their inner product is
\begin{equation}
\langle
{{D^j(g)}^\mu}_\nu,{{D^{j'}(g)}^\rho}_\sigma\rangle=\frac{1}{2j+1}\int_{SU(2)}
\!\!({{D^j}(g)^\mu}_\nu)^*\,{{D^{j'}(g)}^\rho}_\sigma\,{\rm d}_{\rm
H}(g)=
\frac{1}{2j+1}\,\delta^{jj'}\,\delta^{\mu\rho}\,\delta_{\nu\sigma},
\end{equation}
with the Haar measure ${\rm d}_{\rm H}(g)$ on $SU(2)$. $2j+1$ is the
dimension of the representation $(j)$. Taking this over, we have
\begin{equation}
\langle
m,m'\rangle=\frac{1}{2m+1}\,\delta_{m,m'}\hspace{5mm}\mbox{and}\hspace{5mm}
\langle n,n'\rangle=\frac{1}{2n+1}\,\delta_{n,n'},
\end{equation}
thus
\begin{equation}
\langle
m,n|m',n'\rangle=\frac{1}{(2m+1)(2n+1)}\,\delta_{m,m'}\,\delta_{n,n'},
\end{equation}
in spite of the fact that the states appear as mere phase factors.
From the point of view of $U(1)$ this looks quite artificial, but in
fact $U(1)$ is rather an artifact of symmetry reduction.

Applying this to the simple state with $a_{m,n}=1/\sqrt{mn}$ gives
the following contribution of the state $|m,n\rangle$ to the norm
square
\begin{equation}
\frac{a_{m,n}^2}{(2m+1)(2n+1)}=\frac{1}{m(2m+1)n(2n+1)},
\end{equation}
going to $1/4m^2n^2$ for large $m$ and $n$, so in the sense of this
scalar product the state is normalizable.

There is, however, still a problem: $\sqrt{mnk}$, with
$k=\frac{1}{2}(k_++k_-)$ is the volume eigenvalue of a vertex with
holonomy $|m,n\rangle$ with $k_\pm$ denoting the labels of the
adjacent edge functions, which can be chosen arbitrarily without
fluctuation. $k$ describes the cross section of the cell containing
the considered vertex, so $\sqrt{mn/k}$ is the length eigenvalue.
With the inner product borrowed from $SU(2)$ and given edge
holonomies, the length expectation value
$\frac{1}{\sqrt{k}}\langle\sqrt{mn}\rangle$ of a solution is finite,
but the root-mean-square deviation $\Delta\left(\sqrt{mn}\right)$,
containing the expectation value $\langle mn\rangle$, is not: Going
along the $m$ axis with $n=1$, the sum
\begin{equation}
\sum_m\frac{|a_{m,1}|^2\,m}{(2m+1)\cdot3}\sim\sum_m\frac{1}{6m}
\end{equation}
diverges logarithmically, as well as the sum along the $n$ axis. So,
even with the $SU(2)$ inner product, length and volume fluctuations
are infinite. On the other hand, the sums in the direction of the
diagonal and parallel to it, like
\begin{equation}
\sum_m\frac{|a_{m,m}|^2\,m^2}{(2m+1)^2}\sim\sum_m\frac{1}{4m^2},
\end{equation}
are finite.

In conclusion, the Killing conditions in the present model do not
give physically acceptable states, at least the states away from the
diagonal with large wave factors $\ln\frac{n}{m}$ have to be
suppressed, thus violating the Killing conditions. Consider $U_+$:
The solutions are states with $\langle U_+\rangle=0$ without
fluctuations, so it is not a big surprise that something goes wrong
when quantized.

``Infrared divergences" coming from large spins are common in LQG. A
thorough study of divergences of transition amplitudes in covariant
LQG can be found in \cite{Rov}, where they are regulated by recurse
to a spatially finite (visible part of the) universe. Technically
the cutoff is carried out by deforming the gauge group to the
quantum group $SU(2)_q$ with the deformation parameter $q$ being a
function of the cosmological constant $\Lambda$. This leads to a
maximal spin $j_{\rm max}\approx\frac{\pi}{\Lambda\hbar G}$, an
enormous number of order $10^{120}$, but the divergence is only
logarithmical.

Here, on the other hand, we are in canonical LQG and the divergences
concern geometrical quantities like lengths. Further we have
sacrificed the full $SU(2)$ formalism for the sake of simplicity, so
a deformation to the quantum group is not in the scope of our
considerations. (In a calculation of $\hat{U}_+|\psi\rangle=0$ in
the $SU(2)$ framework in \cite{we} divergences of the same kind
occur, which shows that they are not an artifact of this kind of
simplification.)

We just regularize in a different way. An abrupt cutoff at some
maximal length does not look very suitable. A convenient way,
adapted to the form of the results, is to enlarge the powers of $m$
and $n$ in the denominators from $\sqrt{mn}$ to
$(mn)^{\frac{1}{2}+\lambda}$ with a small positive quantity
$\lambda$.

As we do not have a full exact solution, the following calculations
are estimates with the aid of the asymptotic form (\ref{asymp}),
which is decisive for convergence or divergence. We set
\begin{equation}\label{lambda}
a_{m,n}^{(\lambda)}=\frac{\exp\left[\frac{i}{4}\left(b\ln
mn+w\ln\frac{m}{n}\right)\right]}{(mn)^{\frac{1}{2}+\lambda}}.
\end{equation}
\subsection{Length (volume) uncertainties}
The expectation value of the length of a segment containing one
vertex is $\frac{1}{\sqrt{k}}\langle\sqrt{mn}\rangle$, that of
volume is $\sqrt{k}\,\langle\sqrt{mn}\rangle$, with the essential
part
\begin{equation}
\langle\sqrt{mn}\rangle=\sum_{m,n}\frac{|a_{m,n}|^2\sqrt{mn}}{(2m+1)(2n+1)}
\end{equation}
(with normalized coefficients) which is estimated by an integral
over the corresponding continuous asymptotic expressions
\begin{equation}
\langle\sqrt{mn}\rangle\approx\int_1^\infty\frac{{\rm d}m\,{\rm
d}n\,\sqrt{mn}}{(mn)^{1+2\lambda}\,4mn}=\frac{1}{(1-4\lambda)^2}.
\end{equation}

The same estimate for the square of the length leads to
$1/16\lambda^2$ and the lowest-order approximation for $\langle
mn\rangle-\langle\sqrt{mn}\rangle^2$ for small $\lambda$ is
$1/16\lambda^2$, so the mean square root deviations for length and
volume are estimated by
\begin{equation}
\Delta\ell\approx\frac{1}{\sqrt{k}}\frac{1}{4\lambda}\hspace{1cm}\mbox{and}\hspace{1cm}
\Delta V\approx\sqrt{k}\,\frac{1}{4\lambda},
\end{equation}
respectively.

\subsection{Departure from Killing}
Again, in lack of an exact solution for general $b$ and $w$, we take
the modified asymptotic expression for $\alpha(m)$ (\ref{am})
\begin{equation}
\alpha^{(\lambda)}(m)=\frac{\exp\left[\frac{i}{4}(b+w)\ln
m\right]}{m^{\frac{1}{2}+\lambda}}
\end{equation}
and insert it into the left hand side of equation (\ref{73})
\begin{eqnarray}\label{killm}
&&\sqrt{m(m+1)}\;\frac{\exp\left[\frac{i}{4}(b+w)\,\ln
(m+1)\right]}{(m+1)^{\frac{1}{2}+\lambda}}-\sqrt{m(m-1)}\;\frac{\exp\left[\frac{i}{4}(b+w)\,\ln
(m-1)\right]}{(m-1)^{\frac{1}{2}+\lambda}}\nonumber\\
&&-\frac{i}{2}\,(b+w)\,\frac{\exp\left[\frac{i}{4}(b+w)\,\ln
m\right]}{m^{\frac{1}{2}+\lambda}}=:\Delta K(m),
\end{eqnarray}
where the $m$ dependent departure from the Killing property, namely
the difference between space and time derivatives, is denoted by
$\Delta K(m)$. The nonzero asymptotic limit of this expression gives
a measure for the violation of one of the Killing conditions. With
$m\pm1=m(1\pm1/m)$ (\ref{killm}) becomes
\begin{eqnarray}
&&\sqrt{m}\,m^{-\lambda}\,\left(1+\frac{1}{m}\right)^{-\lambda}\exp\left[\frac{i}{4}
(b+w)\left(\ln
m+\ln\left(1+\frac{1}{m}\right)\right)\right]\nonumber\\
&&-\sqrt{m}\,m^{-\lambda}\,\left(1-\frac{1}{m}\right)^{-\lambda}\exp\left[\frac{i}{4}
(b+w)\left(\ln
m+\ln\left(1-\frac{1}{m}\right)\right)\right]\nonumber\\
&&-\frac{i}{2}\,m^{-\frac{1}{2}-\lambda}\:(b+w)\,\exp\left[\frac{i}{4}(b+w)\,\ln
m\right]=\Delta K(m).
\end{eqnarray}
Now we approximate $\left(1\pm\frac{1}{m}\right)^{-\lambda}$ by
$1\mp\frac{\lambda}{m}$, $\ln\left(1\pm\frac{1}{m}\right)$ by
$\pm\frac{1}{m}$ and
$\exp{\left[\pm\frac{i}{4}\frac{b+w}{m}\right]}$ by
$1\pm\frac{i}{4}\frac{b+w}{m}$ and get
\begin{equation}\label{deltaK}
\Delta
K(m)\approx\lambda\left(-2\alpha^{(\lambda)}(m)+O\left(\frac{1}{m}\right)\right).
\end{equation}
Inserting the modified asymptotic limit $\beta^{(\lambda)}(n)$ of
$\beta_n$ in (\ref{UxUy}) into (\ref{74}) gives the same type of
result, a quantity $\Delta K(n)$. The dependence of $\Delta K(m)$ in
(\ref{deltaK}) on $m$ indicates a dispersion of gravitational waves,
in the asymptotic limit the departure goes to the coefficient
$\alpha^{(\lambda)}(m)$ times a constant. A modified state with
coefficients $\alpha^{(\lambda)}(m)$ approaches thus an eigenstate
of $\hat{U}_+$ with eigenvalue $-2\lambda$. As the Killing operators
have the meaning of the difference (time derivative)-(space
derivative), this would mean a constant negative shift of velocity
in the limit of large $m$, namely a speed below the speed of light.
A dependence of $\lambda$ on $b$ and $w$ would be a further source
of dispersion.

The above choice of renormalization was quite ad hoc, like the
mentioned introduction of the quantum group $SU(2)_q$, but
mathematically simpler, suggested just by the asymptotic forms
(\ref{73}) and (\ref{74}). An interesting feature of this result is
that the length/volume uncertainty and $\Delta K$ with its
$m$-dependent parts are in lowest order inverse, i.\,e. that length
and speed appear in this sense complementary, but there is still the
problem of non-commutativity, see below in (5.4.2).

\subsection{Alternative choices}
In the foregoing chapters we had to make some choices, namely in the
factor ordering of the products $K_xE^x$ and $K_yE^y$, the
interpretation of classical derivative terms by differences in the
transition from continuous to discrete and in the renormalization,
necessary to have finite length uncertainties. In lack of guiding
principles in the case of the derivatives ${\cal E}'$ and
$\left(\ln\frac{E^y}{E^x}\right)'${ we have chosen technically
simple options. But it is worthwhile to have also a look at
alternative factor orderings.
\subsubsection{Factor ordering}
In (\ref{+}) the symmetric factor ordering $\sqrt{E}\,K\sqrt{E}$
(with $K$ standing for both $K_x$ and $K_y$ and the analogous
meaning of $E$) for the products $KE$ in $U_+$ was chosen - and
later on tacitly also for $U_-$. The first argument for this choice
is that nonzero expectation values of $U_\pm$ have the meaning of a
difference between space and time derivatives and thus, as physical
quantities, they should be real.

In a more formal language the argument reads also: When the
so-defined operators $U_\pm$ act on holonomies $|m,n\rangle$ with
nonzero $m$ and $n$, they never produce zero-volume states. As
symmetric operators acting on the Hilbert space spanned by a basis
$\{|m,n\rangle\}$ of holonomies with $m$ and $n$ being positive
integers, they are self-adjoint. With the inner product coming from
$U(1)$, the eigenvectors to the eigenvalue zero do not lie in the
Hilbert space, with the $SU(2)$ inner product, they do.

Nevertheless, from the point of view of normalizability, orderings
like $KE$ or even something unnatural as
$E^{-\lambda}K\,E^{1+\lambda}$ would do better. With the choice $KE$
the solution for $U_+|\psi\rangle=0$ with $b=0$, corresponding to
(\ref{58}) and (\ref{pq}) is
\begin{equation}
a_{m,n}=\frac{8\,(m-n)}{(m+n-1)(m+n)(m+n+1)},
\end{equation}
or
\begin{equation}
a_{p,q}=\frac{16\,q}{(2p-1)\,2p\,(2p+1)},
\end{equation}
respectively. With these coefficients the state
$|\psi\rangle=\sum_{m,n}a_{m,n}|m,n\rangle$ is normalizable in the
$U(1)$ norm, whereas the length and volume fluctuation are still
divergent. But normalizability and even finiteness of all
fluctuations of exact solutions of $U_+$ or of $U_-$ is of
relatively little value, as they they do not exactly coincide and
have to be modified anyway for large $m$ and $n$.

\subsubsection{Version (ii) of the operator $\hat{U}_-$}
So far we have worked with version (i), where the derivative of the
wave factor $\ln\frac{E^y}{E^x}$ is replaced by the ``central"
difference and $\hat{U}_+$ and $\hat{U}_-$ commute at every vertex.
So one can find common local solutions and the problem of
non-commutativity would emerge in the construction of common
solutions at neighboring vertices, i.\,e. in the attempt to
construct global solutions.

Version (ii) with its ``left" or ``right" difference (\ref{l+r})
facilitates the construction of global solutions, but here the
problem of non-commutativity pops up locally between the operators
$\hat{U}_\pm$. The situation is qualitatively like that: For each
given vertex holonomy and edge holonomy left (or right) from the
considered vertex there is a solution of $\hat{U}_+|\psi\rangle=0$
and one of $\hat{U}_-|\psi\rangle=0$ (in the following we will
simply say ``a solution of $U_+$ ($U_-$)" for brevity), which have
coefficients $a_{m,n}$ in common, the majority however, does not
coincide.

As the simplest illustration we take a holonomy with $m_-=n_-$ at
the left from the considered vertex $v$ and the ``left" difference,
so that $\left(\ln\frac{E^y}{E^x}\right)'$ gives simply
$\ln\frac{n}{m}$. Further we assume the same edge holonomies with
label $k$ left and right from $v$. In the result, shown in figure 1,
the common coefficients $a_{m,m}$, $a_{m,m\pm1}$ and $a_{m,m\pm2}$
lie inside a diagonal strip in the $(m,n)$ diagram of a vertex. The
nonzero ones (with $m$ and $n$ both odd), indicated by black dots,
are equal to
\begin{equation}\label{amn}
a_{m,n}=\frac{a_{1,1}}{\sqrt{mn}}.
\end{equation}
At positions denoted by {\sf x} the coefficients belonging to a
solution of $\hat{U}_+|\psi\rangle=0$ do not coincide with those
belonging to $\hat{U}_-|\psi\rangle=0$.

\setlength{\unitlength}{0.8mm}

\begin{picture}(130,78)(10,-4)
\put(21,11){\circle*{2}}\put(30,10){0}\put(41,11){\circle*{2}}\put(50,10){\sf
x}
\put(20,20){0}\put(30,20){0}\put(40,20){0}\put(50,20){0}\put(61,21){\sf
x}
\put(21,31){\circle*{2}}\put(30,30){0}\put(41,31){\circle*{2}}\put(50,30){0}
\put(61,31){\circle*{2}} \put(20,41){\sf
x}\put(30,40){0}\put(40,40){0}\put(50,40){0} \put(31,51){\sf
x}\put(41,51){\circle*{2}} \put(45,10){\line(1,1){32}}
\put(20,35){\line(1,1){32}}
\put(20,0){1}\put(30,0){2}\put(40,0){3}\put(50,0){4}
\put(60,1){\vector(1,0){10}} \put(73,0){$m$}
\put(10,10){1}\put(10,20){2}\put(10,30){3}\put(10,40){4}
\put(11,50){\vector(0,1){10}}\put(10,63){$n$}
\put(50,50){\vector(1,1){10}}\put(60,40){\vector(1,1){10}}
\put(96,40){Figure 1: the nonzero common coefficients}
\put(96,32){of the solutions of the two Killing conditi-}
\put(96,24){ons at positions} \put(134,25){\circle*{2}}
\end{picture}

Figure 1 shows the possible common coefficients of solutions of
$U_+$ and $U_-$ at an arbitrary vertex with the holonomy labels $m$
and $n$ along the coordinate axes. They lie inside a strip
$|m-n|\leq2$ along the diagonal. A function whose more off diagonal
coefficients at {\sf x} are either some renormalized mean values of
those of the solutions of $U_+$ and $U_-$, or are simply set equal
to zero, can be considered as a kind of ``approximate solution". The
common coefficients, however, are unique for a given neighboring
vertex holonomy and given holonomies at the adjacent edges. Here
non-commutativity prevents already the existence of local exact
solutions, nevertheless the procedure can be repeated vertex by
vertex to obtain a ``global approximate solution" to the Killing
conditions.

\subsubsection{Wave factor-depending renormalization}
Neither a solution of $U_+$, nor one of $U_-$, constructed under the
above assumptions, would give rise to finite length fluctuations. As
we have seen, for version (ii) of $U_-$ the solutions of $U_+$ and
$U_-$ do not coincide for $|m-n|>2$, for (i) there are common
solutions, but in any case they have to be renormalized, which makes
the search for a common exact solution of the Killing conditions
obsolete. Recall that the contributions to $\langle mn\rangle$ along
diagonal directions are finite (see the behavior of the solutions
displayed in (4.1.2) in $p=\frac{1}{2}(m+n)$). This suggests that
one might renormalize only in the direction of growing $|m-n|$ -
away from the diagonal rather than uniformly in every direction in
$m$ and $n$. In other words: to suppress progressively contributions
with growing wave factor. The most radical method of this kind would
be to cut off everything outside the strip of common coefficients in
figure 1. We choose a modification of the more moderate form. No
renormalization is assumed for zero wave factor, the diagonal
coefficients $a_{m,n}$ in fig. 1 remain untouched, and
renormalization sets in gradually for nonzero wave factors away from
the diagonal.
\begin{equation}\label{renorm1}
a_{m,n}^{(\lambda)}=\frac{a_{m,n}}{|m-n|^\lambda}
\end{equation}
with a small positive $\lambda$, beginning at some distance $|m-n|$
from the diagonal.

To demonstrate this renormalization we take the solution of $U_+$
(\ref{pq}), where the only diagonal term is $a_{1,1}=1$. Here we
reintroduce $a_{1,1}$ as normalizing factor. The expectation value
 $\langle mn\rangle$ is $a_{1,1}/9$ (from the $SU(2)$ inner product)
plus a sum over off-diagonal terms, with all of them, beginning from
$q=1$ ($|m-n|=2$), renormalized in the form (\ref{renorm1}):
\begin{eqnarray}
\langle
mn\rangle&=&a_{1,1}\left[\frac{1}{9}+2\sum_{p=1}^\infty\sum_{q=1}^{p-1}\frac{a_{p,q}^2\,(p^2-q^2)}
{[2(p+q)+1][2(p-q)+1]q^{2\lambda}}\right]\\[2mm]
&<&a_{1,1}\left[\frac{1}{9}+
2\sum_{p=1}^\infty\sum_{q=1}^{p-1}\frac{a_{p,q}^2\,(p^2-q^2)}{2(p+q)2(p-q)q^{2\lambda}}\right]=
a_{1,1}\left[\frac{1}{9}+\frac{1}{2}\sum_{p=1}^\infty
\sum_{q=1}^{p-1}\frac{a_{p,q}^2}{q^{2\lambda}}\right].\nonumber
\end{eqnarray}
Inserting $a_{p,q}$ from (\ref{pq}), denoting the double sum  by
$\Sigma$ and estimating it by integrals gives
\begin{equation}\Sigma\approx\int_1^\infty\!\!\!{\rm
d}p\int_1^{p-1}\!\!\!{\rm
d}q\frac{4q^2(p^2-q^2)}{[(p-1)p(p+1)]^2\,q^{2\lambda}}\approx4\int_1^\infty\!\!\!{\rm
d}p\int_1^{p-1}\!\!\!{\rm d}q\frac{q^2(p^2-q^2)}{p^6\,q^{2\lambda}}.
\end{equation}
In the limit of very small values of $\lambda$ this integral is
equal to $2/15\lambda$, which is much larger than 1/9, so we obtain
the estimate
\begin{equation}
\langle mn\rangle\propto\frac{1}{\lambda}.
\end{equation}
So also this kind of renormalization is sufficient for the
expectation value of (length)$^2$ and the fluctuations of length
become finite.

Inserting $a_{m\pm1,n}^{(\lambda)}$ and $a_{m,n\pm1}^{(\lambda)}$
into the difference equation (\ref{DE}) for $U_+$ with $b=0$, we
obtain in first approximation in the limit of very small $\lambda$
and large $m$ and $n$
\begin{equation}\label{asli}
\Delta
K(m,n)\approx\frac{-2\lambda\,(m-n)\,\sqrt{mn}}{(m+n)^3\,(m-n)^\lambda}=-\frac{\lambda}{4}\,a_{m,n}^{(\lambda)}
\end{equation}
plus higher-order terms in $1/(m+n)$, a result formally analogous to
(\ref{deltaK}). As before in (\ref{deltaK}), also in (\ref{asli})
$\Delta K(m,n)$ is asymptotically proportional to
$a_{m,n}^{(\lambda)}$, so the asymptotic departure from the speed of
light is a negative constant. The difference lies in the asymptotic
limit: In (\ref{deltaK}) this limit means $m\rightarrow\infty$, and
with a modification of the coefficients $\beta(n)$ in (\ref{bn}),
one can derive independently an analogous relation for
$n\rightarrow\infty$, so the asymptotic limit refers to $m$ and $n$
growing symmetrically, i.\,e. to all large length and volume values.

The asymptotic limit in (\ref{asli}), on the other hand, means large
$m+n$, as well as large $|m-n|$. States with small $|m-n|$ are only
slightly modified. Now the physical interpretation is more
plausible: States with small wave factors are only slightly slowed
down, diagonal basic states $|m,m\rangle$ not at all, large wave
factor states are slowed down maximally up to a constant limit.

\section{The Hamiltonian constraint}
\subsection{The operator}
In the Killing operators in the foregoing sections the connection
component $\cal A$ does not occur and for this reason we had to do
only with vertex functions, the edge holonomies could be given
arbitrarily, according to $L$ in the classical prototype
(\ref{1stein}). The dynamics generated by the Hamiltonian
constraint, in contrary, involves all three types of holonomies on
an equal footing, so that the cross section area stands in a
dynamical relation to the vertices.

In accordance with \cite{BD2} we denote the first two terms of the
Hamiltonian constraint in (\ref{constraint}), which contain
connection components, as ``kinetic part" $H_K$, and the rest as
``potential part" $H_P$.The expression for $H$
 contains an inverse volume factor
\begin{equation}
\frac{1}{V}=\frac{1}{\sqrt{{\cal E}E^xE^y}},
\end{equation}
which does not give rise to a densely defined operator in general
and is usually regularized by replacing it by a commutator of
positive powers of connection variables \cite{TT}. But this turns
out as unnecessary, as states with $m$, $n$ or $k$ equal to zero do
not play a role, so that again a piece of the usual mathematical
apparatus of LQG can be avoided.

The Hamiltonian constraint contains first derivatives of $E^x$,
$E^y$ and up to second derivatives of $\cal E$. For this reason we
prefer to associate it to up to four pairs $(v,e)$ of a vertex and
an edge. States at vertex-edge pairs are superpositions of vertex
and edge holonomies in the form
\begin{equation}\label{vertexstate}
|\psi\rangle=\sum_{m,n,k}a_{m,n,k}|m,n,k\rangle.
\end{equation}
$a_{m,n,k}$ is the product of the coefficients $a_{m,n}$ from the
foregoing sections and the probability amplitude of an edge
holonomy.

The potential part is written as
\begin{equation}
H_P=\frac{1}{\sqrt{{\cal E}E^xE^y}}\,P
\end{equation}
with $P$ containing the functions $\cal E$ and ${\cal E}^2$ and the
first and the second $z$-derivatives of $\cal E$. As in the case of
the Killing conditions, derivatives can be replaced by differences
in different ways, either containing the variables $m$, $n$, and $k$
at the considered vertex and edge, or only the labels from
neighboring ones. In the following, in the first case we write
$P(m,n,k)$, in the second one only $P$, and assume pairs $(v,e)$
with the edge right from the vertex. As a concrete choice, in
addition to $m$, $n$, $k$, belonging to $(v,e)$, the quantity
$P(m,n,k)$ may be chosen to involve the labels from the vertex left
to $v\equiv v_i$ and those of the two edges left from $e$, indexed
by $i-1$ and $i-2$, respectively:
\begin{eqnarray}\label{Pi}
&&P(m,n,k)=\frac{1}{4}\,(k-k_{i-1})^2+\frac{1}{4}\left[k\left(\ln\frac{n}{m}
-\ln\frac{n_{i-1}}{m_{i-1}}\right)\right]^2\\[2mm]
&&\hspace{20mm}-\frac{1}{2}\,k\,(k-k_{i-1})\left[\ln(nm)-\ln(n_{i-1}m_{i-1})\right]+
k\,(k+k_{i-2}-2k_{i-1}).\nonumber
\end{eqnarray}
(There are of course other possible choices.) In any case $P(m,n,k)$
results in a diagonal operator, and for the difference equation
below it does not make a difference, whatever interpretation of the
differences is chosen.

With the above choice, for every state on a sequence
``edge-vertex-edge" left from $v$,
\begin{equation}\label{initial}
|k_{i-2}\rangle\otimes|m_{i-1},n_{i-1},k_{i-1}\rangle
\end{equation}
(the vertex function $|m_{i-2},n_{i-2}\rangle$ does not contribute,
as there is no second derivative in $E^x$, $E^y$) one can formulate
and solve a difference equation for the pair $(v,e)$, and so on,
going step by step from one vertex-edge pair to the next one, from
the left to the right. In general, there will be a superposition of
states (\ref{initial}) at the chosen sequence with the probability
amplitudes
\begin{equation}\label{amp}
\sum_{\{m,n,k\}}a_{m-2,\,n-2,\,k-2}\cdot
a_{m_{i-1},\,n_{i-1},\,k_{i-1}},
\end{equation}
where $\{m,n,k\}$ refers to the labels at the two vertices and the
two edges left from $v$, and the obtained states at $(v,e)$ have to
be multiplied with (\ref{amp}) for every state of the form
(\ref{initial}). And so on from left to right\ldots$\;$ Starting
from such an ``initial-value" state at some sequence
``edge-vertex-edge", one can of course, do everything also from
right to left for pairs $(e,v)$ with the edge left from the vertex.

For the kinetic part $H_{K}$ (with $\kappa=1$ and the inverse volume
factor included) we choose again the ordering
$K_aE^a\rightarrow\sqrt{E^a}\,K_a\,\sqrt{E^a}$ and write the inverse
volume term to the left, so we obtain the factor ordering
``connection left from triad" (and $\kappa=1$)
\begin{equation}\label{ord}
H_K=\frac{1}{\sqrt{\cal
E}}\,K_x\sqrt{E^x}\,K_y\sqrt{E^y}-\frac{1}{\sqrt{E^y}}\,K_x\sqrt{E^x}\,{\cal
A}\,\sqrt{\cal E}-\frac{1}{\sqrt{E^x}}\,K_y\sqrt{E^y}\,{\cal
A}\,\sqrt{\cal E}.
\end{equation}
The action of the corresponding operator on a basic state
$|m,n,k\rangle$ is the following
\begin{eqnarray}\label{HK}
&&H_K|m,n,k\rangle=\frac{1}{4}
\left[\sqrt{\frac{mn}{k}}\left(|m+1,n+1,k\rangle-|m-1,n+1,k\rangle\right.\right.\nonumber\\
&&\hspace{40mm}-\left.|m+1,n-1,k\rangle+|m-1,n-1,k\rangle\right)\nonumber\\[2mm]
&&\hspace{20mm}+\sqrt{\frac{mk}{n}}\left(|m+1,n,k+1\rangle-|m-1,n,k+1\rangle\right.\\
&&\hspace{40mm}-\left.|m+1,n,k-1\rangle+|m-1,n,k-1\rangle\right)\nonumber\\[2mm]
&&\hspace{20mm}+\sqrt{\frac{nk}{m}}\left(|m,n+1,k+1\rangle-|m,n-1,k+1\rangle\right.\nonumber\\[-2mm]
&&\hspace{40mm}-\left.\left.|m,n+1,k-1\rangle+|m,n-1,k-1\rangle\right)\rule{0mm}{5mm}\right].\nonumber
\end{eqnarray}

Zero $m$, $n$, or $k$: For zero $k$, for example, the second and the
third prefactor simply vanish, in the first term, for $k=0$ we have
only states of the form $|m\pm1,n\pm1,0\rangle$ on the right-hand
side. So this term leaves the $k=0$ states among themselves, as well
as $H_P$, being diagonal, does not mix states with $n=0$ and $m=0$
with other ones. The second kinetic term acts analogously to the
first one on $|m\rangle\otimes|k\rangle$ and the third one on
$|n\rangle\otimes|k\rangle$. Due to this property we can
consistently avoid zero-volume states. Furthermore, $H_K$ does not
lead to relations between states with positive and negative labels
and we can restrict ourselves to $m,n,k>0$. For this reason we also
do not need absolute values in (\ref{HK}).

\subsection{The difference equation}

When the Hamilton operator is imposed to annihilate physical states
of the form (\ref{vertexstate}),
\begin{equation}
\hat{H}_K|\psi\rangle=-\frac{\hat{P}}{\sqrt{{\cal
E}E^xE^y}}\,|\psi\rangle,
\end{equation}
for each holonomy $|m,n,k\rangle$ on the right-hand side, the
inverse volume factor gives $1/\sqrt{mnk}$, as $P$ is a diagonal
operator. After multiplication with $4\sqrt{mnk}$, we obtain the
Hamiltonian constraint difference equation
\begin{eqnarray}\label{diffequation}
&&\left[\sqrt{mn}\left(\sqrt{(m-1)(n-1)}\,a_{m-1,n-1,k}-
\sqrt{(m+1)(n-1)}\,a_{m+1,n-1,k}\right.\right.\nonumber\\
&&\hspace{12mm}\left.-\sqrt{(m-1)(n+1)}\,a_{m-1,n+1,k}+
\sqrt{(m+1)(n+1)}\,a_{m+1,n+1,k}\right)\nonumber\\[2mm]
&&+\sqrt{mk}\left(\sqrt{(m-1)(k-1)}\,a_{m-1,n,k-1}-
\sqrt{(m+1)(k-1)}\,a_{m+1,n,k-1}\right.\nonumber\\
&&\hspace{12mm}-\left.\sqrt{(m-1)(k+1)}\,a_{m-1,n,k+1}+
\sqrt{(m+1)(k+1)}\,a_{m+1,n,k+1}\right)\\[2mm]
&&+\sqrt{nk}\left(\sqrt{(n-1)(k-1)}\,a_{m,n-1,k-1}-
\sqrt{(n+1)(k-1)}\,a_{m,n+1,k-1}\right.\nonumber\\
&&\hspace{12mm}-\left.\left.\sqrt{(n-1)(k+1)}\,a_{m,n-1,k+1}+
\sqrt{(n+1)(k+1)}\,a_{m,n+1,k+1}\right)\right]\nonumber\\[2mm]
&&+4P(m,n,k)\,a_{m,n,k}=0.\nonumber
\end{eqnarray}

Connecting in general 13 coefficients to each other, this equation
looks quite complicated. Lets begin with $(m,n,k)=(1,1,1)$:
\begin{equation}\label{3}
a_{1,2,2}+a_{2,1,2}+a_{2,2,1}=-2\,P(1,1,1)\,a_{1,1,1},
\end{equation}
the next simplest case, with $(m,n,k)=(1,1,2)$, $(1,2,1)$, or
$(2,1,1)$, leads to the three relations
\begin{eqnarray}\label{4}
&&\sqrt{3}\,a_{1,2,3}+\sqrt{3}\,a_{2,1,3}+a_{2,2,2}=a_{1,2,1}+a_{2,1,1}
-2\,P(1,1,2)\,a_{1,1,2},\nonumber\\
&&\sqrt{3}\,a_{1,3,2}+\sqrt{3}\,a_{2,3,1}+a_{2,2,2}=a_{1,1,2}+a_{2,1,1}
-2\,P(1,2,1)\,a_{1,2,1},\\
&&\sqrt{3}\,a_{3,1,2}+\sqrt{3}\,a_{3,2,1}+a_{2,2,2}=a_{1,1,2}+a_{1,2,1}
-2\,P(2,1,1)\,a_{2,1,1}.\nonumber
\end{eqnarray}
For a systematic look at the difference equation it is convenient to
order the coefficients according to the sum
\begin{equation}
p=m+n+k.
\end{equation}
Equation (\ref{3}) relates 3 of the 6 coefficients with $p=5$ to an
initial value $a_{1,1,1}$ (with $p=3$), the three equations
(\ref{4}) relate 7 of the 10 coefficients with $p=6$ on the
left-hand side to three initial values $a_{1,1,2}$, $a_{1,2,1}$ and
$a_{2,1,1}$ with $p=4$ on the right-hand side.

In this way, when we visualize the coefficients in a 3-dimensional
grid with axes $m$, $n$, and $k$, (see figure 2) the coefficients
$a_{m,1,1}$, $a_{1,n,1}$, and $a_{1,1,k}$ on the axes are free
initial or boundary values for the solutions, but they do not
determine them uniquely. Graphically for each $p$ the coefficients
are represented by points on a triangle, spanned by $(p-2,1,1)$,
$(1,p-2,1)$, and $(1,1,p-2)$ in the $(m,n,k)$ coordinate system.

In the next step the three coefficients $a_{1,2,2}$, $a_{2,1,2}$,
and $a_{2,2,1}$ in (\ref{3}) plus the free coefficients $a_{1,1,3}$,
$a_{1,3,1}$, and $a_{3,1,1}$ with $p=5$ are related to 12 of 15
coefficients with $p=7$, written on the left-hand side in the
following way
\begin{eqnarray}
&&\sqrt{3}\,a_{2,2,3}+\sqrt{3}\,a_{2,3,2}+3\,a_{1,3,3}=a_{2,1,2}+a_{2,2,1}+
\sqrt{3}\,a_{1,3,1}+\sqrt{3}\,a_{1,1,3}\nonumber\\
&&\hspace{55mm}-a_{1,1,1}-2\,P(1,2,2)\,a_{1,2,2},\nonumber\\[3mm]
&&\sqrt{3}\,a_{2,2,3}+\sqrt{3}\,a_{3,2,2}+3\,a_{3,1,3}=a_{1,2,2}+a_{2,2,1}+
\sqrt{3}\,a_{1,1,3}+\sqrt{3}\,a_{3,1,1}\nonumber\\
&&\hspace{55mm}-a_{1,1,1}-2\,P(2,1,2)\,a_{2,1,2},\nonumber\\[3mm]
&&\sqrt{3}\,a_{2,3,2}+\sqrt{3}\,a_{3,2,2}+3\,a_{3,3,1}=a_{1,2,1}+a_{2,1,2}+
\sqrt{3}\,a_{1,3,1}+\sqrt{3}\,a_{3,1,1}\\
&&\hspace{55mm}-a_{1,1,1}-2\,P(2,2,1)\,a_{2,2,1},\nonumber\\[3mm]
&&\sqrt{6}\,a_{1,2,4}+\sqrt{6}\,a_{2,1,4}+a_{2,2,3}=\sqrt{3}\,a_{1,2,2}+\sqrt{3}\,a_{2,1,2}
-2\,P(1,1,3)\,a_{1,1,3},\nonumber\\[3mm]
&&\sqrt{6}\,a_{1,4,2}+\sqrt{6}\,a_{2,4,1}+a_{2,3,2}=\sqrt{3}\,a_{1,2,2}+\sqrt{3}\,a_{2,2,1}
-2\,P(1,3,1)\,a_{1,3,1}\nonumber\\[3mm]
&&\sqrt{6}\,a_{4,1,2}+\sqrt{6}\,a_{4,2,1}+a_{3,2,2}=\sqrt{3}\,a_{2,1,2}+\sqrt{3}\,a_{2,2,1}
-2\,P(3,1,1)\,a_{3,1,1}.\nonumber
\end{eqnarray}
Having calculated these coefficients for certain holonomies of the
form (\ref{initial}) at the left neighboring vertex and the two
edges left from $(v,e)$, for a general state (\ref{vertexstate}) at
$(v_{i-1},e_{i-1})$ and $(v_{i-2},e_{i-2})$, one has to multiply the
results by the product (\ref{amp}) and form superpositions.

\subsection{Number of free coefficients}
In each further step, functions at $e_{i-1}$ and at $(v,e)$ give
rise to a solution at $(v_{i+1},e_{i+1})$, but do not determine it
uniquely, as we have seen above, at every step there remain
undetermined coefficients. In each further step $\frac{p(p+1)}{2}-3$
coefficients with $m+n+k=p$ are related to the coefficients with
$m+n+k=p-2$, where 3 of the former ones remain undetermined and
serve as initial values for the next group with $m+n+k=p+2$.
Generally, for $m+n+k=p$ we have
\begin{equation}
\frac{1}{2}(p-2)(p-1)
\end{equation}
coefficients. Inserting all combinations with $m+n+k=p-2$ into the
difference equation (\ref{diffequation}) we obtain
$\frac{1}{2}(p-4)(p-3)$ linear equations, relating the coefficients
with $p$ to those with $p-2$ and $p-4$. Calculating coefficients
going step by step from $p$ to $p+2$, we can construct special
solutions, either with only even, or with only odd $p$'s. At every
step with a certain value of $p$ we have
\begin{equation}
\frac{1}{2}\,(p-3)(p-4)
\end{equation}
coefficients determined by the Hamiltonian constraint and
\begin{equation}\label{free}
\frac{1}{2}\,(p-1)(p-2)-\frac{1}{2}\,(p-3)(p-4)=2p-5
\end{equation}
free ones.

Summing up the number of coefficients with even and odd $m+n+k=3$
from 3 to $p$, we have
\begin{equation}\label{total}
\frac{1}{6}\,p\,(p-1)(p-2)
\end{equation}
coefficients, where
\begin{equation}
(p-2)^2
\end{equation}
of them can be chosen freely, so that
\begin{equation}\label{best.H}
\frac{1}{6}\,(p-2)(p-3)(p-4)
\end{equation}
are determined.

\setlength{\unitlength}{1mm}

\begin{picture}(80,90)(10,5)
\put(30,40){\line(0,1){50}} \put(30,40){\line(1,0){50}}
\put(30,40){\line(-2,-3){20}} \put(22,28){\line(5,2){30}}
\put(14,16){\line(5,2){60}} \put(51,40){\circle*{3}}
\put(30,60){\circle*{3}} \put(21.5,28){\line(1,4){8}}
\put(22,28){\circle*{3}}\put(51,40){\line(-1,1){21}}
\put(13.5,15.5){\line(1,4){16}}\put(72,40){\line(-1,1){41}}
\put(30,40){\circle*{3}}\put(14,16.5){\circle*{3}}
\put(72,40){\circle*{3}}\put(30,81){\circle*{3}}
\put(43,28){\circle*{3}}\put(51.1,60.1){\circle*{3}}
\put(22,50){\circle*{3}} \put(13,10){$m$} \put(78,36){$n$}
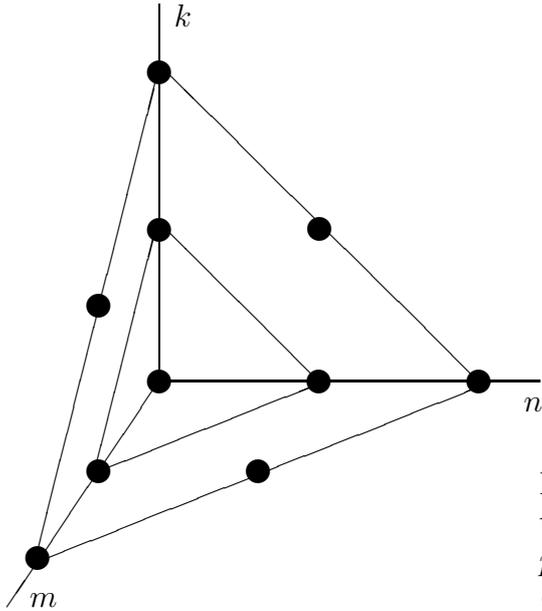
\put(32,87){$k$} \put(80,25){Figure 2: The coefficients for $p=3$
(at $(1,1,1)$,} \put(80,20){taken here as origin of the axes),}
\put(80,15){$p=4$ (smaller triangle), and $p=5$} \put(80,10){(larger
triangle) in $(m,n,k)$ coordinates.}
\end{picture}

Had we chosen a different factor ordering, e.\,g. ``connection left
from triad" including the inverse volume factor, as it is often
advocated for the Hamiltonian constraint operator, the square
root-coefficients of the $a_{m,n,k}$ in (\ref{diffequation}) would
become different, but the structure of the system of equations would
remain the same, particularly the numbers of the free and the
determined $a_{m,n,k}$'s.

As mentioned before, in the above considerations the diagonal
functions $P(m,n,k)$ are chosen in a different way than in
\cite{BD2}. There the operator acts individually on every vertex,
with a dependence between vertex functions mediated only via edge
labels, and create new vertices. For the choices advocated here,
they depend also directly on the neighboring functions and do not
change the graph.

\subsection{Asymptotic behavior}
To estimate the asymptotic behavior of solutions for large $m$, $n$,
and $k$ we replace the coefficients $a_{m,n,k}$ by differentiable
functions $a(m,n,k)$ and approximate the differences in
(\ref{diffequation}) by partial derivatives. The resulting
differential equation is separable, when the derivatives of $\ln
(E^y/E^x)$ and $\ln(E^xE^y)$ in $P(m,n,k)$ are formulated in terms
of only neighboring labels like, for example, in (\ref{central}).
So, for an easy study of the asymptotic behavior we choose such a
form of $P$, which does not affect the results obtained so far,
$P(m,n,k)$ is simply replaced by $P$ in the above formulae. This is
sufficient for separability. Further, it is possible, and suitable
for simplifying the equation, and for a more uniform asymptotic
behavior in $m$, $n$, and $k$, to reformulate $P$ completely in
terms of derivatives,
\begin{eqnarray}\label{reform}
&&P=\frac{1}{2}\left({\cal E}^2\right)''-\frac{3}{4}\left({\cal
E}'\right)^2-\frac{1}{4}({\cal E}')^2(\ln E^xE^y)'\\
&&\hspace{10mm}+\frac{1}{4}\left({\cal
E}^2\,\ln\frac{E^y}{E^x}\right)'\left(\ln\frac{E^y}{E^x}\right)'
-\frac{1}{8}\left({\cal
E}^2\right)'\left[\left(\ln\frac{E^y}{E^x}\right)^2\right]'.\nonumber
\end{eqnarray}
If we now formulate also $\left({\cal E}^2\right)'$ and $\left({\cal
E}^2\right)''$ in terms of neighboring edge labels,
\begin{equation}
\left({\cal E}^2\right)'\rightarrow k_{i+1}^2-k_{i-1}^2,
\hspace{1cm} \left({\cal E}\right)''\rightarrow
k_{i+2}+k_{i-2}-k_{i+1}-k_{i-1}
\end{equation}
to the left and the right (or on one side), also $k$ disappears from
$P$ and the asymptotic differential equation becomes
\begin{eqnarray}
&&\sqrt{mn}\frac{\partial^2}{\partial m\,\partial
n}\left(\sqrt{mn}\,a(m,n,k)\right)+
\sqrt{mk}\frac{\partial^2}{\partial m\,\partial
k}\left(\sqrt{mk}\,a(m,n,k)\right)\nonumber\\
&&+\sqrt{nk}\frac{\partial^2}{\partial n\,\partial
k}\left(\sqrt{nk}\,a(m,n,k)\right)=-P\,a(m,n,k).
\end{eqnarray}
With the product ansatz $a(m,n,k)=M(m)\,N(n)\,K(k)$ this becomes
\begin{eqnarray}
&&\sqrt{mn}\frac{\rm d}{{\rm d}m}\left(\sqrt{m}\,M\right)\frac{\rm
d}{{\rm d}n}\left(\sqrt{n}\,N\right)K+\sqrt{mk}\frac{\rm d}{{\rm
d}m}\left(\sqrt{m}\,M\right)N\frac{\rm d}{{\rm d}
k}\left(\sqrt{k}\,K\right)\nonumber\\
&&+\sqrt{nk}\,M\frac{\rm d}{{\rm d}
n}\left(\sqrt{n}\,N\right)\frac{\rm d}{{\rm d}
k}\left(\sqrt{k}\,K\right)=-P\,M\,N\,K.
\end{eqnarray}
Introducing ${\cal M}=\sqrt{m}\,M$, ${\cal N}=\sqrt{n}\,N$, and
${\cal K}=\sqrt{k}\,K$, the derivatives written in form of primes,
and dividing by $M\,N\,K$ gives
\begin{equation}\label{cal}
\sqrt{mn}\,\frac{{\cal M}'}{M}\frac{{\cal
N}'}{N}+\sqrt{mk}\,\frac{{\cal M}'}{M}\frac{{\cal K}'}{K}
+\sqrt{nk}\,\frac{{\cal N}'}{N}\frac{{\cal K}'}{K}=-P,
\end{equation}
finally
\begin{equation}\label{149}
\left(m\frac{{\cal M}'}{\cal M}\right)\left(n\frac{{\cal N}'}{\cal
N}\right)+\left(m\frac{{\cal M}'}{\cal M}\right) \left(k\frac{{\cal
K}'}{\cal K}\right)+\left(n\frac{{\cal N}'}{\cal
N}\right)\left(k\frac{{\cal K}'}{\cal K}\right)=-P.
\end{equation}

When $P=0$, e.\,g. all derivatives in (\ref{H}) are zero, the
simplest solution to (\ref{149}) is given by $\frac{{\cal M}'}{\cal
M}=\frac{{\cal N}'}{\cal N}=\frac{{\cal K}'}{\cal K}=0$, which leads
to
\begin{equation}
a_{m,n,k}\rightarrow\frac{\rm const.}{\sqrt{mnk}}.
\end{equation}
When in (\ref{H}) only the derivatives of $\cal E$ (or the
differences of $k$ in (\ref{Pi})) are zero,
\begin{equation}
P=\frac{1}{4}\left[{\cal
E}\left(\ln\frac{E^y}{E^x}\right)'\right]^2, \hspace{1cm}\mbox{or}
\hspace{1cm}
P=\frac{1}{4}\left[k\,\left(\ln\frac{n}{m}-\ln\frac{n_{i-1}}{m_{i-1}}\right)\right]^2,
\end{equation}
respectively, is positive. Then for states, where $m\frac{{\cal
M}'}{\cal M}$, $n\frac{{\cal N}'}{\cal N}$, and $k\frac{{\cal
K}'}{\cal K}$ are (approximately) equal, these expressions are
imaginary, as the r.\,h. side of (\ref{149}) is negative then.

More generally, in standard macroscopic situations the cross section
area is much larger than its derivative, i.\,e. ${\cal E}\gg{\cal
E}'$ (particularly in the linear approximation for small waves the
cross section area $\cal E$ is constant along $z$), so that
the positive definite term 
in (\ref{Pi}), coming from ${\cal E}^2\left((\ln
(E^y/E^x))'\right)^2$, will be dominant, so that $P(m,n,k)>0$ and
also $P>0$.

With this in mind, we separate off the part dependent on $m$ and set
both sides equal to an (imaginary) constant
\begin{equation}
m\frac{{\cal M}'}{\cal M}=-\frac{P+n\frac{{\cal N}'}{\cal
N}\:k\frac{{\cal K}'}{\cal K}}{n\frac{{\cal N}'}{\cal N}+
k\frac{{\cal K}'}{\cal K}}=i\mu.
\end{equation}
For $\cal M$ we find a function proportional to $m^{i\mu}$, which
means that
\begin{equation}\label{mu}
M(m)\propto\frac{1}{\sqrt{m}}\:e^{i\mu\ln m}.
\end{equation}
Continuing separation gives analogously
\begin{equation}\label{nu}
N(n)\propto\frac{1}{\sqrt{n}}\:e^{i\nu\ln
n}\hspace{1cm}\mbox{and}\hspace{1cm}
K(k)\propto\frac{1}{\sqrt{k}}\:e^{i\rho\ln k}
\end{equation}
with
\begin{equation}
\mu\nu+\mu\rho+\nu\rho=P.
\end{equation}
(The equation is separable also for the choice (\ref{Pi}), where $P$
is a linear function of $k$. Without the reformulation
(\ref{reform}), $P$ would be a quadratic function in $k$ and in the
exponent in $K(k)$ also an (imaginary) quadratic function in $k$
would appear, making $K(k)$ oscillate much faster in $k$ than $M$
and $N$ in $m$ and $n$.)

For the chosen factor ordering in (\ref{ord}) the factors
$\frac{1}{\sqrt{m}}$ and $\frac{1}{\sqrt{n}}$ in (\ref{mu}) and
(\ref{nu}) match with the corresponding factors (\ref{am}) and
(\ref{bn}) in the unrenormalized solutions of the Killing
conditions. This would not be the case, had we chosen a symmetric
factor ordering in $H_K$. In the case of $m$, $n$, and $k$ growing
at the same order of magnitude, a symmetric choice for the
separation constants w.\,r. to the three kinds of holonomies,
$\mu=\nu=\rho=\sqrt{P/3}$, will occur. Then also the oscillation
frequencies of $M$, $N$, and $K$ are the same and the asymptotic
behavior of the coefficients
of such a 
state 
is
\begin{equation}\label{amnk}
a_{m,n,k}\rightarrow\frac{1}{\sqrt{mnk}}\,e^{i\sqrt{\frac{P}{3}}\,\ln(mnk)}.
\end{equation}

With the inner product induced from $SU(2)$ the solutions become
normalizable, although with the same problem of diverging
length/volume uncertainties as the solutions of the Killing
conditions. An ad hoc renormalization of the form
\begin{equation}
\frac{1}{\sqrt{mnk}}\rightarrow\frac{1}{(mnk)^{\frac{1}{2}+\lambda}},
\end{equation}
in analogy to (\ref{lambda}), which solves the divergence problem,
results in a state that does not exactly solve the Hamiltonian
constraint. If we trust the considered model, this might mean that
in quantum gravity there are small energy fluctuations as in other
QFTs, or that gravity has to be necessarily coupled to something
``coming from outside". This also alters the relation between the
Hamiltonian and the diffeo constraint, and leads also to departures
from a constant speed of gravitational waves, in accordance with the
results of Sections 4 and 5.

\subsection{Flat space} Flat space, as a state without waves in
either direction, is classically given by $U_+=U_-=0$, see
(\ref{Up},\ref{Um}), with both signs of the derivative terms.
Classically this means $K_xE^x=K_yE^y=0$ as well as $b=w=0$
(background and wave factors are constant). In quantum theory the
common solution of $\hat U_+|\psi\rangle=\hat U_-|\psi\rangle=0$
((\ref{73}) and (\ref{74}) with $b=w=0$ and $a_{1,1}=1$) is simply
\begin{equation}\label{vak}
a_{m,n}=\frac{1}{\sqrt{mn}},
\end{equation}
nonzero only for odd $m$ and $n$. The conditions $b=w=0$ have no
quantum analog in the Killing conditions, and the edge functions
remain undetermined.

A (renormalized) state of the form describes one-dimensional quantum
length fluctuations in the $z$ direction in absence of cross section
fluctuations and with one and the same holonomy at two neighboring
vertices left and right of the considered vertex.

A solution of the Hamiltonian constraint with $P=0$ has the form
\begin{equation}\label{vak1}
a_{m,n,k}=\frac{1}{\sqrt{mnk}}.
\end{equation}
Here the probability amplitude for an edge state $|k\rangle$ is also
equal to $1/\sqrt{k}$. So a three-dimensional local approximative
state for flat space is obtained, a renormalized version with
\begin{equation}\label{vak2}
a_{m,n,k}\propto\frac{1}{(mnk)^{\frac{1}{2}+\lambda}}
\end{equation}
gives a picture of 3 dimensional fluctuations in flat space. $P=0$
would mean that all derivatives in $P$ add up to zero. This is
certainly not true in most of the cases, so the above model does not
display all details of flat space in the given framework, but in a
fluctuating, macroscopically homogeneous vacuum state at least the
statistical mean values of the derivatives, and thus also of $P$,
are zero, so that (\ref{vak1}) or (\ref{vak2}) have a meaning as a
good statistical approximation.

\section{Summary}
Thanks to its effective one-dimensionality, the symmetry-reduced
model considered in this paper facilitates investigations about
possible dispersion of gravitational waves. In the quantum regime
the two independent Killing equations, guaranteeing dispersion-free
propagation, act as conditions on physical states. The relative
simplicity of the model does not require too much mathematical
sophistication and in all of this paper we tried generally to keep
the mathematical apparatus as modest as possible, in favor of
explicit calculations. So, in comparison with \cite{Hos}, the issue
of the inverse volume can be circumvented. As the crucial issues
happen in the asymptotic limit for large volume, nothing
qualitatively different would happen, had we reformulated the
inverse volume as in \cite{TT}.

The holonomies being effectively mere phase factors, we have chosen
holonomy operators looking formally like $U(1)$ operators. In
reward, the difference equations become simpler, allowing even for
explicit solutions. On the other hand, concerning the derivatives of
$\cal E$ and $\ln\frac{E^y}{E^x}$, we stuck to the classical
prototype and represented first/second derivatives by first/second
differences. In this pronounced non-local approach this involves
necessarily more than one edge, or vertex, respectively, of a
one-dimensional SNW, and nonzero Poisson brackets of local
expressions containing derivatives are reflected in form of the
non-commutativity of operators in the discrete regime - at
neighboring vertices for version (i) of $U_-$ with central
differences, - and at one and the same vertex for version (ii) with
one-sided difference. Note that $U_\pm$ are not symmetry-generating
constraints of the theory, they just pick out certain states, so we
do not have to take their commutators into account as new
conditions, which would lead to an inconsistent infinite tower of
conditions. Our interpretations of derivatives make the construction
of either local solutions (for version (i)) or of only approximative
global solutions (for version (ii)) possible.

The obtained solutions are not normalizable in the obvious $U(1)$
inner product, and even with the inner product borrowed from
$SU(2)$, the true gauge group of quantum gravity, which falls off
more rapidly for large spins, length uncertainties diverge. Problems
of this kind are usually overcome by continuous wave packets. If the
quantities $b$ and $w$ in $U_\pm$ and the potential part $P$ in $H$
were continuous, one could construct well-behaved wave packets
without invoking anything ``from outside", but they are discrete.
The discrete interpretation in $U_-$ is ambiguous and a matter of
choice; in $U_+$ it is at least very natural, but also here the
normalization problem of solutions is present. Particularly the
difference $k_+-k_-$ in (\ref{49}) is integer, so superpositions of
asymptotic solutions are periodic.

This makes some renormalization necessary. In Section 5 we have
shown two ad-hoc examples that manage to suppress sufficiently the
influence of large geometric eigenvalues $m$ and $n$. By modifying
solutions, we commit violations of the Killing conditions, i.\,e. a
departure from an exactly constant propagation speed of
gravitational waves. An estimation shows that this speed and the
metric quantities appear as mutually complementary - a typical
situation in quantum theory. In comparison, the second kind of
renormalization introduced in (5.4.2), which fits to the diagonal
domain of coincidence of solutions of $U_\pm$ in fig. 1, is more
selective, it suppresses large wave factors. When the considered
kinds of renormalization are ascribed to the interaction with matter
fields, both the Hamiltonian and the diffeo constraint would have to
be combined with the corresponding matter quantities.

To make it easier to follow the lines of argumentation, we give a
brief overview of the locations of the different steps in the main
part:\\[2mm]
(4.1.2) solution of $\hat{U}_+|\psi\rangle=0$ for the r.\,h. side
$b=0$,\\[1mm]
(4.1.3) non-normalizability for the $U(1)$ inner product,\\[1mm]
(4.2.1) $\hat{U}_-$, central difference for
$\left(\ln\frac{E^y}{E^x}\right)'$,\\[1mm]
(4.2.2) common local solution for
$\hat{U}_+|\psi\rangle=\hat{U}_-|\psi\rangle=0$, not normalizable in
$U(1)$,\\[1mm]
(5.1) inner product from $SU(2)$ $\rightarrow$ normalizable
solutions, but divergent length and volume fluctuations,
renormalization,\\[1mm]
(5.2) finite fluctuations,\\[1mm]
(5.3) corresponding departure from the Killing conditions,
expectation value,\\[1mm]
(5.4.1) alternatives of the factor ordering of $\hat{U}_\pm$ and of
the discrete counterparts of derivatives,\\[1mm]
(5.4.2) one-sided difference in $\hat{U}_-$, non-commutativity
shifted from neighboring vertices to single vertices,\\[1mm]
(5.4.3) alternative renormalization.\\[2mm]
The conclusion of the last points in (5.4.2) is that the
compatibility of the solutions of $\hat{U}_+$ and those of
$\hat{U}_-$ is very restricted, namely to coefficients with
$|m-n|\leq2$. It may be considered as a sort of lucky interplay
between non-commutativity and the need of renormalization that it is
sufficient to renormalize the coefficients that do not coincide, as
done in (5.4.3).

The first renormalization attempt leads to slowing down of
gravitational waves by a constant amount in the apparently classical
limit of asymptotically large geometric quantities $m$ and $n$ - a
rather untrustworthy result. One may however observe that this
method, using the factor $(mn)^{-\lambda}$, would correspond to an
inner product that falls off slightly quicker than the $SU(2)$
induced one. If one can find an argument for such an inner product,
renormalization can be avoided and the remaining issue is that of
non-commutativity. One could try to find asymptotically coincident
solutions of $U_\pm$ and modify them for small values of $m$ and
$n$. In this way waves characterized by macroscopic large length and
volume values would propagate undisturbed at constant speed $c$.

The second method leads to the physically much more plausible result
of a constant slowing down in the limit of large wave factors, and
an undisturbed background with zero wave factor. As a consequence of
dispersion, in any case the notions of wave and background factor
retain only an approximate meaning.

In Section 6 we have applied the described methods to the
Hamiltonian constraint in a non-graph changing version. Representing
derivatives by differences including variables coming either from
left or from the right of a considered vertex or edge, as before, is
a choice that disregards the generally assumed commutativity of the
operators at neighboring vertices. It is traded for a faithful
mapping of derivatives to corresponding differences and thus avoids
ultra-locality. With such an approach involving only one-sided
differences, it is possible to construct global solutions going step
by step from vertex to vertex and edge to edge. What is sacrificed
is the off-shell commutativity on all of the space of unconstrained
(kinematical) states, what remains is on-shell commutativity, which
is trivial on states with $\hat{H}|\psi\rangle=0$. As a reward we
gained quantum solutions for the Hamiltonian constraint as well as
for the Killing conditions. Reflecting the degree of derivatives by
the corresponding degree of differences yields a stronger bond
between quantities at neighboring vertices. As can be seen from the
(infinite) number of free parameters in (\ref{free}), there is a
large solution space.

Interestingly, the applied methods lead to the same normalizability
problem for the Killing conditions and for the Hamiltonian
constraint, illustrated by the same asymptotic behavior. This
indicates a close relationship between gravitational wave dispersion
and slight departures from the Hamiltonian constraint, and also the
exact equivalence of the Hamiltonian and the diffeo constraint gets
perturbed. As the Hamiltonian constraint describes the symmetry of
pure gravity under changes of the local units of time, a
modification indicates the presence of some other fields, acting as
a regulator for quantum gravity states. Some small fluctuation
around the expectation value zero of the Hamiltonian would be a
common feature with other QFTs.

As further possibilities, infrared divergences of large geometric
quantities (= spins in full LQG) would in principle be cut off in a
finite universe, or one could at least argue with the relation
between the diameter of the visible universe and the Planck length,
as it is done in \cite{Rov}. At first sight this looks quite inept
and ridiculous, but numerically the logarithm of this ratio (recall
that the length uncertainty diverges logarithmically) is of the
order of $10^2$, so the root mean square deviation of length and
volume is of order 10. This is much, if the expectation value
amounts to a few Planck lengths, but for macroscopic lengths it
becomes reasonably tiny, but even in view of the not completely
absurd numerics it is hard to accept.

In \cite{Rov} the cosmological constant appears generally as
regulator for LQG, this might avoid invoking matter fields. More
recently there are also attempts to promote the Immirzi parameter to
a scalar field \cite{Immirzi}, or to add topological terms as fields
to the action of GR, see \cite{Giacomo} and citations therein,
related directly, but not minimally coupled, to gravity.

Even if we do not have a physical principle as guideline for
renormalization, the main conclusion that can be drawn from our
results from a {\em fully quantum} system is that in the scope of
the present model some dispersion of gravitational waves,
by whatever physical reasons, and even with a ``better" inner product,
has to be expected.\\

\noindent{\bf Acknowledgement:} The author thanks Seth Major for
valuable help and discussions.

\end{document}